\documentclass[showpacs, aps, superscriptaddress, twocolumn]{revtex4-2}
\usepackage{mathrsfs}
\usepackage{array}
\usepackage{amsmath}
\DeclareMathOperator{\csch}{csch}
\usepackage{mathtools}
\newcommand\norder[1]{\vcentcolon\mathrel{#1}\vcentcolon}
\usepackage[ngerman,english]{babel}
\usepackage{stackengine}
\usepackage{scalerel}
\usepackage{amssymb}
\usepackage{graphicx}
\usepackage{color,xcolor}
\usepackage{booktabs}
\usepackage{amstext}
\DeclareMathOperator{\Tr}{Tr}
\usepackage{amsfonts}
\usepackage{float}
\usepackage{bm}
\usepackage{epstopdf}
\usepackage{color}
\usepackage{tcolorbox}

\begin{document}

\title{The strong-coupling quantum thermodynamics of quantum Brownian motion \\ 
based on the exact solution of its reduced density matrix}
\author{Chuan-Zhe Yao}
\affiliation{Department of Physics and Center for Quantum
Information Science, National Cheng Kung University, Tainan 70101,
Taiwan}
\author{Wei-Min Zhang}
\email{wzhang@mail.ncku.edu.tw}
\affiliation{Department of Physics
and Center for Quantum Information Science, National Cheng Kung
University, Tainan 70101, Taiwan}

\begin{abstract}
We derive the quantum thermodynamics of quantum Brownian motion from the exact solution of its reduced density matrix. We start from the total equilibrium thermal state between the Brownian particle and its reservoir, and solve analytically and exactly the reduced density matrix of the system by taking the partial trace over all the reservoir states.  We find that the reduced Hamiltonian and the reduced partition function of the Brownian motion must be renormalized significantly, as shown in the general nonperturbative renormalization theory of quantum thermodynamics for open quantum systems we developed recently [Phys.~Rev.~Res.~\textbf{4}, 023141 (2022)]. 
The reduced Hamiltonian contains not only a frequency shift but also a squeezing pairing interaction, where
a momentum-dependent potential is generated naturally from the strong coupling between the Brownian particle and the reservoir, after traced over all the reservoir states. The resulting exact reduced density matrix of the Brownian motion is given by a squeezing thermal state. Moreover, beyond the weak coupling limit, in order to obtain correctly the reduced partition function of the Brownian motion, one must take into account the non-negligible changes of the reservoir state induced by the system-reservoir coupling. Using the exact solutions of the reduced density matrix, the reduced Hamiltonian as well as the reduced partition function of the Brownian motion, we show that the controversial results obtained from the different definitions of internal energy and the issue of the negative heat capacity in the previous studies of strong-coupling quantum thermodynamics are resolved. 
\end{abstract}

\maketitle
\section{Introduction}
\label{intro}
The study of quantum thermodynamics beyond the weak-coupling limit has received tremendous attentions in recent years. Under the strong coupling between the system and its reservoir, many new thermodynamic phenomena may occurs.
The foundation of thermodynamic laws may also be challenged. For example, it has been realized recently that quantum features, such as quantum coherence and quantum entanglement, could  enhance the energy conversion efficiency in nanoscale heating systems \cite{engine1,engine2,engine3,engine4,engine5,engine6,engine7,WM2}. This has significant implications for designing and optimizing quantum heat engines in the future. More importantly, the study of quantum thermodynamics under strong coupling provides reinterpretation and new understanding of thermodynamic laws. The traditional thermodynamic laws may require modifications or extensions to accommodate the quantum features in the nano- and atomic-scale systems \cite{WM1,WM3,QT1,QT2,QT3,QT4,QT5,QT6,Seifert16,Jarzynski17,seclaw1,seclaw2,seclaw3,heat,PH07,CB2022,entangle1,entangle2,entangle3,PH06,PH08,PH09,nega4,nega5,HC2,PH20,MF1,MF2,MF3,MF4,MF5,MF6}. This motivates researchers to attempt to establish a general framework for quantum thermodynamics at the strong coupling, to discover new quantum thermodynamic phenomena, and to explore the foundation of thermodynamics \cite{QT1,QT2,MF1,MF3,Seifert16,Jarzynski17,PH20,WM1}. The quantum heat capacity at strong coupling also becomes an important issue because it may be used to justify the validity of the third law of thermodynamics \cite{PH06,PH08,PH09,nega4,nega5,HC2} and may also have inherently connection with quantum entanglement \cite{entangle1,entangle2,entangle3}.

However, due to the strong coupling, many thermodynamic quantities, including the internal energy and the heat capacity, have to be redefined. Some contradictory results on heat capacity in strong-coupling quantum Brownian motion were found \cite{PH06,PH08,PH09}, which are arisen from the inconsistent definitions of the internal energy. In order to overcome the difficulties in finding a useful theory for quantum thermodynamics  at  strong coupling, an effective Hamiltonian known as the Hamiltonian of mean force was introduced and widely used \cite{PH20,MF1,MF2,MF3,MF4,MF5,MF6,Seifert16,Jarzynski17}. In such approach, the thermal state of the reservoir is proposed to be invariant because of its macroscopic nature. 
Nevertheless, as shown in our recent work \cite{WM1}, for any system coupling to a thermal reservoir with the strong system-reservoir coupling, both the system and the reservoir undergo a non-equilibrium dynamical evolution. The final equilibrium temperature of the system-reservoir state must be renormalized \cite{WM1}. This indicates that the strong coupling between the system and reservoirs can result in a non-negligible change for both the system and reservoir states, which naturally questions the validity of the Hamiltonian of mean force in the strong-coupling regime.

Furthermore, we also find that the system Hamiltonian should be renormalized in the way that the strong coupling effect between the system and the reservoir must be correctly incorporated after traced over all the reservoir states. In a recent work \cite{YW2022}, we have derived analytically the exact master equation for a generalized quantum Brownian motion. It goes beyond the previous derivations of the master equation for the quantum Brown motion by Caldeira and Leggett for a Ohmic spectral density \cite{Leggett1983} and by Hu, Paz and Zhang for the general color noise \cite{HPZ1992}. In our recent work \cite{YW2022}, all the physical quantities have been renormalized with the standard framework of many-body nonequilibrium dynamics \cite{WM1}. It shows that the renormalization not only modifies the reduced system Hamiltonian, but also induces a momentum-dependent potential to the Brownian motion. Such a new potential induced by the system-reservoir coupling was not recognized in both the Caldeira-Leggett master equation and the Hu-Paz-Zhang master equation \cite{Leggett1983,HPZ1992}. 
It is also omitted further in the recent investigations to the strong-coupling quantum thermodynamics \cite{PH20,HC2,Grabert1,MF2,MF3,MF4,MF5,MF6}, even in cases where some of them did not employ the approach of the Hamiltonian of mean force.

On the other hand, in some other studies \cite{QBMME1,QBMME2}, including our recent work \cite{YW2022}, it has been pointed out that renormalization effects stemming from the system-reservoir coupling can lead to the squeezing phenomenon for quantum Brownian motion. The effects of quantum squeezing in quantum thermodynamics remain an evolving field of study to date. Recent research has unveiled novel thermodynamic behaviors with quantum squeezing \cite{squeezed thermal1,squeezed thermal2,squeezed thermal3,squeezed thermal4,squeezed thermal5,squeezed thermal6}. The squeezing phenomena can affect the thermodynamic quantities, such as thermal population, internal energy, and thermodynamic work. Nevertheless, in previous studies of the quantum thermodynamics for quantum Brownian motion \cite{PH06,PH08,PH09,PH20,nega4,nega5,HC2,Grabert1,MF1,MF2,MF3,MF4,MF5,MF6}, there is a lack of discussion on the influence of the renormalization-induced squeezing effect  on the thermodynamics of  Brownian motion. In fact, we find that if the  system Hamiltonian is correctly renormalized \cite{YW2022}, the resulting quantum Brownian motion, when it reaches the thermal equilibrium with its reservoir, is characterized by a typical squeezing thermal state. A detailed discussion of this matter will be presented in the next section, where we will also show the relation between the momentum-dependent potential and the renormalization-induced squeezing property in the quantum Brownian motion.

After all, in this paper, we study analytically the quantum Brownian particle interacting weak or strongly with its reservoir, 
where both the Brownian particle and all the particles in the reservoir are modeled as harmonic oscillators that are 
linearly coupled to each other \cite{Feynman1963, Leggett1983}. We begin with a total thermal equilibrium state in 
which the Brownian particle and the reservoir particles are entangled completely, which can be easily realized 
under the quantum dynamical evolution for an arbitrary state of the Brownian particle coupling to a thermal reservoir \cite{WM1}. 
The total thermal equilibrium state is usually called a partition-free state in the literature, which is indeed a Gibbs state of 
the total system (Brownian particle plus its reservoir). 
We then use the coherent-state integral approach and group theory to carry out the partial trace over all the reservoir states without using any assumption or approximation. From this analytical solution, we find how the state of Brownian motion is intricately changed by the reservoir, and how a momentum-dependent potential (a squeezed pairing effect) is naturally generated from the linear coupling between the Brownian particle and the reservoir.
Furthermore, we utilize the faithful matrix representation in group theory \cite{Gilmore,group,bookw} to deal with the operator ordering problem in the reduced density matrix  
 and obtain the  reduced Hamiltonian, the corresponding reduced density matrix, as well as the reduced partition function of the Brownian motion. Such a rigorous derivation has not been carried out in the literature, due to the difficulty in dealing with the infinite degrees of freedom in the reservoir. It shows that the reduced density matrix of the Brownian motion can be also expressed as a Gibbs state with the reduced Hamiltonian containing inevitably a momentum-dependent potential that is generated from the system-reservoir coupling, which has been omitted in the previous studies \cite{PH06,PH08,PH20,HC2,Grabert1,MF1,MF2,MF3,MF4,MF5,MF6}. 

By further diagonalizing the reduced Hamiltonian through a Bogoliubov transformation, we show that the Bogoliubov quasiparticle 
 is subjected to an effective harmonic oscillator potential in a new position-momentum coordinate system. As an result, we
obtain an extended Bose-Einstein distribution for Brownian motion with pairing interactions, which describes the relations between the
 occupation and squeezing as well as the renormalized frequency and pairing strength. As a self-consistent 
check, we compute the heat capacity from the different definitions based on the  reduced Hamiltonian and reduced partition function, respectively, and find the consistent solution. In order to compare our results with the controversial results found in the literature, we also numerically calculate the heat capacity based on the incomplete-reduced Hamiltonian and incomplete-reduced partition function of the Brownian motion that studied in Refs.~\cite{PH06,PH08,PH09,nega4,nega5}. We show how  contradictory results were generated from the improper definitions of internal energy and partition function in these previous works for the strong-coupling quantum thermodynamics. 

The rest of the paper is 
organized as follows. In Sec.~\ref{sec2}, we solve the reduced density matrix and the corresponding reduced system Hamiltonian 
based on our nonperturbation renormalization theory for quantum thermodynamics \cite{WM1}. The analytical analysis  and numerical results  
of thermodynamic quantities for quantum Brownian motion are presented in Sec.~\ref{sec3}. A conclusion is given in Sec.~\ref{concl}. The detailed derivations 
of the formulas are presented in Appendices.

\section{The exact reduced density matrix of the Brownian motion}
\label{sec2}
The quantum Brownian motion describes the behavior of a quantum particle interacting with a thermal bath. It is modeled by the Hamiltonian
\cite{Feynman1963, Leggett1983},
\begin{align}
H_{\rm tot}=&H_\textsc{S}+H_\textsc{E}+H_\textsc{SE}\notag\\
=&\dfrac{P^2}{2m}+\dfrac{1}{2}M\omega_0^2Q^2+\sum\limits_k\dfrac{f_k^2Q^2}{2m_k\omega_k^2}\notag\\
&+\sum\limits_k\left(\dfrac{p_k^2}{2m_k}+\dfrac{1}{2}m_k\omega_k^2x_k^2\right)-\sum\limits_kf_kQx_k, \label{H1}
\end{align}
in which the Brownian particle $(P,Q)$ is considered to be trapped in a harmonic potential, and the last term in $H_\textsc{S}$ represents the counter-term to remove the possible unphysical divergence of the renormalized frequency-shift. The reservoir is modeled as a collection of an infinite number of harmonic oscillators $(p_k, x_k)$ with a continuous frequency distribution. The system-reservoir interaction is described by the linear coupling $\{ f_k \}$ between the Brownian particle position $Q$ and all the reservoir mode positions $\{ x_k \} $, depicted by the last term in the above Hamiltonian.

For the convenient calculations in the coherent state basis \cite{group,bookw}, we rewrite the above Hamiltonian in the particle number representation,
\begin{align}
\label{H2}
H_{\rm tot}=&\hbar\omega_\textsc{S} a^\dagger a+\sum\limits_k\hbar\omega_k b_k^\dagger b_k\notag\\
&+\sum\limits_k\hbar V_k\left(a^\dagger b_k+b_k^\dagger a+a^\dagger b_k^\dagger+b_k a\right),
\end{align}
where $a^\dagger$ and $b_k^\dagger$ ($a$ and $b_k$) are the bosonic creation (annihilation) operators for the system and the reservoir with energy quanta $\hbar\omega_\textsc{S}$ and $\hbar\omega_k$, respectively. The effective frequency of the system incorporating the counter-term  is given by $\omega_\textsc{S}^2=\omega_0^2+\sum_kf_k^2/Mm_k\omega_k^2$. The coupling parameter $V_k=-f_k/2\sqrt{Mm_k\omega_S\omega_k}$ is the coupling amplitude between the system and the k-mode in the reservoir, which can be weak or strong.
The form of Eq.~(\ref{H2}), which is exactly the same as Eq.~(\ref{H1}), shows that the linear coupling between the system and the reservoir contains the particle pair production and annihilation, which will modify significantly the renormalized system Hamiltonian when the coupling becomes strong, as we will see later.

To study the thermodynamics of quantum Brownian motion, we consider the total system to be in a thermal equilibrium state. Such
a state can always be realized when the system and the reservoir reach their steady state, even if they begin with a decoupled initial state, i.e., the system can be in an arbitrary initial state and the reservoir is initially in a thermal reservoir \cite{WM1}. Then the steady-state density matrix of the total system (the Brownian particle plus its reservoir) is given by 
\begin{align}
\rho_{\rm tot}=\frac{1}{Z_{\rm tot}}e^{-\beta H_{\rm tot}},  \label{tGS}
\end{align} 
where $\beta=1/k_\textsc{B} T$, and $T$ is the equilibrium temperature of the total system at the steady state. Note that the final equilibrium temperature at the steady state can be significantly different from the initial equilibrium temperature of the reservoir in strong coupling, as shown explicitly in \cite{WM1}, and Eq.~(\ref{tGS}) is a highly entangled state between the system and its reservoir. 

The key issue to derive the quantum thermodynamics of the Brownian motion lies on the state of the Brownian motion itself, which is completely described by the reduced density matrix $\rho_S$. The reduced density matrix $\rho_S$ is determined by the partial trace over all the reservoir states from the total density matrix $\rho_{\rm tot}$, 
\begin{align}
\rho_\textsc{S}=\Tr_\textsc{E} [\rho_{\rm tot}] = \frac{1}{Z_{\rm tot}} \Tr_\textsc{E} [e^{-\beta H_{\rm tot}}],
\end{align}
which is usually a very difficult task.  Fortunately, for the quantum Brownian motion, 
the total density matrix is a Gaussian-type state so that the partial trace is exactly doable with the help of group theory and coherent state method \cite{group} or other reliable methods. In the coherent state representation, the total density matrix can be expressed as
\begin{align}
\langle\boldsymbol{z}|\rho_{\rm tot}|\boldsymbol{z^\prime}\rangle\!=\!\dfrac{\sqrt{\det\bm{\Omega}}}{Z_{\rm tot} }\exp\!\left[\dfrac{1}{2}\begin{pmatrix}
\boldsymbol{z^\dagger} & \boldsymbol{z^{\prime T}}\end{pmatrix}\!\!
\begin{pmatrix}
\boldsymbol{\Omega}  & \boldsymbol{\Pi} \\
\boldsymbol{\Pi^*}  &\boldsymbol{\Omega^*} 
\end{pmatrix}\!\!
\begin{pmatrix}
\boldsymbol{z^\prime}\\
\boldsymbol{z^*}
\end{pmatrix}\!
\right],
\label{rho_tot}
\end{align}
where $|\bm{z}\rangle=|z_S,\bm{z_E}\rangle=\exp\{ z_S a^\dag +\sum_k z_k b^\dag_k \} |0 \rangle $ is
the unnormalized coherent state which is also the eigenstate of the annihilation operators $(a, \{ b_k \})$ with complex eigenvalue $\bm{z} = (z_S,\bm{z}_E)$.  The normalized factor of the coherent states has been moved into the integral measure of the coherent state identity resolution for the convenience of calculations, as one will see later. The factor $\sqrt{\det\bm{\Omega}}$ comes from the normal ordering of the operators in coherent state representation \cite{norder1,norder2}, that is, $\langle z|A(a^\dag,a)|z'\rangle=\norder{A(z^*,z')}$. 
More explicitly, the constant arises from the simple relation $\langle z|e^{\kappa(a^\dag a+aa^\dag)}|z'\rangle=e^\kappa \exp[{e^\kappa(z^*z'+z'z^*)}]$ based on the group theory \cite{group}, 
see more details in Eq.~(\ref{A9}).
The matrix block $\boldsymbol{\Omega}$ is a Hermitian matrix and $\boldsymbol{\Pi}$ is an symmetric matrix, both are $(k_c+1)\times (k_c+1)$ matrices, 
\begin{align}
&\boldsymbol{\Pi}=\begin{pmatrix}
\Pi_{SS}&\bm{\Pi}_{S\bm{E}}\\
\bm{\Pi}_{S\bm{E}}^*&\bm{\Pi_{EE}}
\end{pmatrix},\
\boldsymbol{\Omega}=\begin{pmatrix}
\Omega_{SS}&\bm{\Omega}_{S\bm{E}}\\
\bm{\Omega}_{S\bm{E}}^{*}&\bm{\Omega_{EE}}
\end{pmatrix}
\end{align}
where $k_c$ is the total number of the oscillation modes in the reservoir and is infinitely large in principle. With the aid of the faithful matrix representation of a subgroup of the Lie group $Sp(2k_c+4)$ \cite{group} (see appendix \ref{appendixA} for the detail derivation), they are given by
\begin{subequations}
\label{Omega_Pi}
\begin{align}
\bm{\widetilde{\Omega}} &=\bigg\{\! \begin{pmatrix} \bm{0} & \bm{I} \end{pmatrix}
\exp \bigg[\begin{pmatrix}
\bm{D}&\bm{R} , \\
-\bm{\hat{R}}&-\bm{\widetilde{D}}
\end{pmatrix}
\bigg]
\begin{pmatrix}
\bm{0}\\ \bm{I}
\end{pmatrix} \! \bigg\}^{-1} \\
\bm{\Pi} & =\bigg\{ \! \begin{pmatrix} \bm{I}&\bm{0} \end{pmatrix}
\exp \bigg[  \begin{pmatrix} \bm{D}&\bm{R} \\ -\bm{\hat{R}}&-\bm{\widetilde{D}} \end{pmatrix} \bigg] 
\begin{pmatrix} \bm{0}\\ \bm{I} \end{pmatrix} \! \bigg\}
\bm{\widetilde{\Omega}},
\end{align}
\end{subequations}
where $\widetilde{\bm D}$ denotes a reflection in the minor diagonal of $\bm{D}$, and $\hat{\bm R}$ is a reflection in the major diagonal of the  $\bm{R}$. The matrices $\bm{D}$ and $\bm{R}$ are determined by the parameters in the total Hamiltonian $H_{\rm tot}$ of Eq.~(\ref{H2}),
\begin{align}
\bm{D}=-\dfrac{\hbar\beta}{2}\begin{pmatrix}
\omega_S&\bm{V}_{S\bm{E}}\\
\bm{V}_{S\bm{E}}^\dag&\boldsymbol{\omega_{E}}
\end{pmatrix},\
\bm{R}=-\dfrac{\hbar\beta}{2}\begin{pmatrix}
\bm{\bar{V}}_{S\bm{E}}&0\\
\bm{0}&\bm{V}_{S\bm{E}}^\dag
\end{pmatrix},
\label{M}
\end{align}
in which $\bm{\bar{V}}_{S\bm{E}}$ is the inversion of $\bm{V}_{S\bm{E}}$. More explicitly,
\begin{subequations}
\begin{align}
&\boldsymbol{\omega_E}=\!\begin{pmatrix}
&\omega_{k_1} & & & \\[-2mm]
& & \omega_{k_2}  &  & \hspace{-2mm}\mbox{\Huge 0}\\
& \hspace{-2mm}\mbox{\Huge 0} & & & \hspace{-2mm}\ddots 
\end{pmatrix} \!, \  
\bm{V}_{S\bm{E}}=\begin{pmatrix}
V_{k_1} & V_{k_2} & \cdots 
\end{pmatrix}, \\
& \boldsymbol{\widetilde{\omega}_E}=\!\begin{pmatrix}
&\ddots & & &\hspace{-2mm}\mbox{\Huge 0} \\
& & &\hspace{-2mm} \omega_{k_2}  &\\[-2mm]
&\hspace{-2mm}\mbox{\Huge 0} & & &\omega_{k_1} 
\end{pmatrix} \!, \
\bm{\bar{V}}_{S\bm{E}}=\begin{pmatrix}
\cdots& V_{k_2} & V_{k_1}
\end{pmatrix}.
\end{align}
\end{subequations}

Taking the partial  trace over all the reservoir states via the Gaussian integral (See Appendix\ref{appendixA}), 
the reduced density matrix of the quantum Brownian motion in the coherent state representation can be obtained analytically,
\begin{align}
\label{coherent_S}
\langle z_S|{\rho_S}
&|z^\prime_S\rangle=
\int \! d\mu(\boldsymbol{z_E})\langle z_S,\boldsymbol{z_E}|
\rho_{tot} 
|z_S^\prime,\boldsymbol{z_E}\rangle\notag\\
=&\dfrac{1}{Z^r_{S_0}}\exp\left[\dfrac{1}{2}\begin{pmatrix}
z_S^\dagger & z_S^{\prime T}
\end{pmatrix}\!
\begin{pmatrix}
\Omega_{S}  & \Pi_{S} \\
\Pi_{S}^*  & \Omega_{S}
\end{pmatrix}\!
\begin{pmatrix}
z_{S}^\prime\\
z_S^*
\end{pmatrix}
\right] 
\end{align}
Here the coherent state integral measure $d\mu(\bm{z_E})=\prod_k \frac{dz_kdz^*_k}{2\pi i}e^{-|z_k|^2}$ and the factor $e^{-|z_k|^2}$
comes from the normalization of the coherent state. The resulting matrix after the integral is found to be
\begin{align}
\begin{pmatrix}
\Omega_{S}  & \Pi_{S} \\
\Pi_{S}^*  & \Omega_{S} 
\end{pmatrix}&=\begin{pmatrix}
\Omega_{SS}  & \Pi_{SS} \\
\Pi_{SS}^*  & \Omega_{SS}
\end{pmatrix}
+\!\begin{pmatrix}
\bm{\Omega}_{S\bm{E}}  & \bm{\Pi}_{S\bm{E}} \\
\bm{\Pi}^*_{S\bm{E}}  & \bm{\Omega}^*_{S\bm{E}} 
\end{pmatrix}\!\notag\\
&\times\left[\boldsymbol{I}\!-\!\begin{pmatrix}
\boldsymbol{\Omega_{EE}}  & \boldsymbol{\Pi_{EE}} \\
\boldsymbol{\Pi_{EE}^*}  & \boldsymbol{\Omega^*_{EE}} 
\end{pmatrix}\!\right]^{-1}
\!\!\begin{pmatrix}
\bm{\Omega}_{\bm{E}S}  & \bm{\Pi}_{\bm{E}S} \\
\bm{\Pi}^*_{\bm{E}S}  & \bm{\Omega}^*_{\bm{E}S} 
\end{pmatrix}\! .
\label{OPS_matrix}
\end{align}
The partition function in Eq.~(\ref{coherent_S}) is given by
\begin{align}
Z^r_{S_0}=&\dfrac{Z_{\rm tot}}{\sqrt{\det\bm{\Omega}}}\left\Vert\bm{1}-\begin{pmatrix}
\bm{\Omega_{EE}}&\bm{\Pi_{EE}}\\
\bm{\Pi^*_{EE}}&\bm{\Omega^*_{EE}}
\end{pmatrix}
\right\Vert^{1/2}\notag\\
=&\sqrt{\dfrac{1}{(1-\Omega_S)^2-|\Pi_S|^2}}.
\label{ZSR0}
\end{align}

In general, the thermal reservoir is significantly larger compared to the system. This means that the total number of oscillation modes $k_c$ in the reservoir is infinite, and so does the dimension of $\bm{\Omega}$ and $\bm{\Pi}$. In this situation, it is very difficult to directly find the renormalization of the reduced system encompassing all the reservoir effect through the matrix operations in Eq.~(\ref{OPS_matrix}). Therefore, we also apply the imaginary-time path integral approach in the coherent state representation (See Appendix \ref{appendixB}) to calculate the partial trace over the reservoir states that is characterized by a continuous spectral density $J(\epsilon)=2\pi\sum_k|V_k|^2\delta(\epsilon-\epsilon_k)$. After integrating out all the degree of freedom of the reservoir, we have
\begin{align}
&\dfrac{d}{d\tau}\left.\begin{pmatrix}
\Omega_S(\tau)&\Pi_S(\tau)\\
\Pi_S^*(\tau)&\Omega_S^*(\tau)
\end{pmatrix}\right \vert_{\tau\rightarrow\hbar\beta} \!\!\!\! =\!\begin{pmatrix}\!
-\epsilon_S&0\\0&\epsilon_S
\!\end{pmatrix}  \begin{pmatrix}
\Omega_S(\tau)&\Pi_S(\tau)\\
\Pi_S^*(\tau)&\Omega_S^*(\tau)
\end{pmatrix} \notag\\
&+\!\int_0^{\hbar\beta}\!\! d\tau'\bigg\{\! \begin{pmatrix}
g(\hbar\beta-\tau')&g(\hbar\beta-\tau')\\
g(\tau'-\hbar\beta)&g(\tau'-\hbar\beta)
\end{pmatrix}\!\begin{pmatrix}
\Omega_S(\tau')&\Pi_S(\tau')\\
\Pi_S^*(\tau')&\Omega_S^*(\tau')
\end{pmatrix}\notag\\[+1mm]
&+\![g'(\tau'\!-\!\hbar\beta)\!+\!g'(\hbar\beta\!-\!\tau')]\!\begin{pmatrix}
1&\!1\\-\!1&\!-\!1
\end{pmatrix}
\!\begin{pmatrix}
\Omega_S(\tau')&\Pi_S(\tau')\\
\Pi_S^*(\tau')&\Omega_S^*(\tau')
\end{pmatrix}\!\!\bigg\}
\label{OPS_dot}
\end{align}
subjected to the initial conditions $\Omega_S(0)=1$ and $\Pi_S(0)=0$, where the integral kernels $g(\tau)$ and $g'(\tau)$ are determined by the spectral density $J(\omega)=2\pi\sum_k|V_k|^2\delta(\omega-\omega_k)$,
\begin{subequations}
\label{g}
\begin{align}
g(\tau)&=\int\dfrac{d\omega}{2\pi}J(\omega)e^{-\omega\tau} ,\\
g'(\tau)&=\int\dfrac{d\omega}{2\pi}J(\omega)e^{-\omega\tau}/(e^{\beta\hbar\omega}-1).
\end{align}
\end{subequations}
Note that the integro-differential equation Eq.~(\ref{OPS_dot}) is solvable only in the limit $\tau\rightarrow\hbar\beta$ at which the periodic boundary condition along the imaginary time $\tau$ can be satisfied (See Appendix \ref{appendixB} for details).
Thus, the solutions of $\Omega_S(\tau)$ and $\Pi_S(\tau)$ at the given temperature $\tau=\hbar\beta$ are given by the following inverse Laplace transformation
\begin{widetext}
\begin{align}
\label{OPS_Laplace}
\begin{pmatrix}
\Omega_S(\hbar\beta)&\Pi_S(\hbar\beta)\\
\Pi_S^*(\hbar\beta)&\Omega_S^*(\hbar\beta)
\end{pmatrix}=\mathcal{L}^{-1}\left[\begin{pmatrix}
s+\omega_S+\Sigma(s)+\Sigma'(s)+\Sigma'(-s)&\Sigma(s)+\Sigma'(s)+\Sigma'(-s)\\
\Sigma(-s)-\Sigma'(s)-\Sigma'(-s)&s-\omega_S+\Sigma(-s)-\Sigma'(s)-\Sigma'(-s)
\end{pmatrix}^{-1}\right],
\end{align}
\end{widetext}
where $\Sigma(s)=\mathcal{L}[g(\tau)]$ and $\Sigma'(s)=\mathcal{L}[g'(\tau)]$ are the Laplace transform of the integral kernels.

Now, using the same technique of Gaussian integrals, we can find explicitly the particle number occupation $n=\Tr[a^\dagger a \rho_S]$ and the squeezing parameter $s =\Tr[ aa \rho_S]$ in terms of the matrix element $\Omega_S $ and $\Pi_S $,
\begin{subequations}
\label{NS}
\begin{align}
&n=\dfrac{\Omega_S(1-\Omega_S)+|\Pi_S|^2}{(1-\Omega_S)^2-|\Pi_S|^2} , \\
&s=\dfrac{\Pi_S}{(1-\Omega_S)^2-|\Pi_S|^2}.
\end{align}
\end{subequations}
Alternatively, the matrix elements $\Omega_S $ and $\Pi_S $ in Eq.~(\ref{coherent_S}) can be expressed in terms of the occupation $n$ and the squeezing $s$ which are physically measurable quantities, 
\begin{subequations}
\begin{align}
&\Omega_S =1-\dfrac{1+n}{(1+n)^2-|s|^2} , \\
&\Pi_S =\dfrac{s}{(1+n)^2-|s|^2}.
\end{align}
\end{subequations}
Furthermore, by applying the faithful matrix representation of $H_6$, a subgroup of $Sp(4)$ \cite{group}, the reduced density matrix $\rho_S$ 
of Eq.~(\ref{coherent_S}) can be solved explicitly (see Appendix A),
\begin{align}
\label{rho_S}
\rho_S  &\!=\!\dfrac{1}{Z_S^r}e^{\alpha a^{\dagger2}}e^{\gamma(a^\dagger a+aa^\dagger)}e^{\alpha^* a^2} \notag \\
& \!=\! \dfrac{1}{Z^r_S}\exp\left[\eta(a^\dagger a+aa^\dagger)+\delta a^{\dagger2}+\delta^* a^2\right],
\end{align}
where the reduced partition function $Z_S^r$ and the coefficients $\alpha$, $\gamma$, and $\eta$, $\delta$ in the reduced density matrix $\rho_S$ of Eq.~(\ref{rho_S}) can be expressed in terms of the particle number occupation $n$ and the squeezing parameter $s$ as following:
\begin{subequations}
\label{12}
\begin{align}
&Z^r_S=\sqrt{\dfrac{\Omega_S}{(1-\Omega_S)^2-|\Pi_S|^2}} =\sqrt{n^2+n-|s|^2}  \label{eZSR} ,\\
&\alpha=\dfrac{\Pi_S}{2}=\dfrac{s}{2[(1+n)^2-|s|^2]}\\
&\gamma=\ln\sqrt{\Omega_S}=\ln\sqrt{1-\dfrac{1+n}{(1+n)^2-|s|^2}}\\
&\eta=\dfrac{n+\frac{1}{2}}{2\sqrt{\left\vert(n+\frac{1}{2})^2-|s|^2\right\vert}}\ln\left[\dfrac{\sqrt{\left\vert(n+\frac{1}{2})^2-|s|^2\right\vert}-\frac{1}{2}}{\sqrt{\left\vert(n+\frac{1}{2})^2-|s|^2\right\vert}+\frac{1}{2}}\right]\\
&\delta=\dfrac{-s}{2\sqrt{\left\vert(n+\frac{1}{2})^2-|s|^2\right\vert}}\ln\left[\dfrac{\sqrt{\left\vert(n+\frac{1}{2})^2-|s|^2\right\vert}-\frac{1}{2}}{\sqrt{\left\vert(n+\frac{1}{2})^2-|s|^2\right\vert}+\frac{1}{2}}\right]\label{21c} .
\end{align}
\end{subequations}
Thus, the reduced density matrix of the Brownian motion is completely solved from the total thermal equilibrium state of the system coupled with its reservoir. It shows that when the system-reservoir coupling effect is fully included, the quantum Brownian motion in the equilibrium state is a typical squeezing thermal state.  Based on the above exact solution of the reduced density matrix,
quantum thermodynamics of the Brownian motion can be unambiguously studied.

In particular, the reduced partition function of the Brownian particle in Eq.~(\ref{rho_S}) is also given by $Z^r_S=\sqrt{\Omega_S}Z^r_{S_0}$ (see Appendix A). Using Eq.~(\ref{ZSR0}), we find the relation between the reduced partition function $Z^r_S$ and the total partition function $Z_{\rm tot}$ as following:
\begin{align}
\label{ZSR}
    Z^r_{S} 
      & = Z_{\rm tot}\sqrt{\dfrac{\Omega_S}{\det\bm{\Omega}}}\left\Vert\bm{1}-\begin{pmatrix}
\bm{\Omega_{EE}}&\bm{\Pi_{EE}}\\
\bm{\Pi^*_{EE}}&\bm{\Omega^*_{EE}}
\end{pmatrix}
\right\Vert^{1/2}.
\end{align}
As shown in Eq.~(\ref{Omega_Pi}),  Eq.~(\ref{M}) and Eq.~(\ref{OPS_matrix}), the matrix element $\bm{\Omega}$ as well as $\bm{\Omega_{EE}}$ and $\bm{\Pi_{EE}}$ are not only related to the reservoir mode frequencies $\bm{\omega_E}$, but also depend on the system frequency $\omega_S$ and the system-reservoir coupling $\bm{V}_{S\bm{E}}$. In other words, Eq.~(\ref{ZSR}) implies the fact that 
the state of reservoir cannot remain unchanged. 
Instead, the reservoir state must also have the corresponding changes induced by the system-reservoir coupling, in particular when the coupling becomes strong.
As a consequence,  
\begin{align}
Z^r_S\neq&\frac{Z_{\rm tot}}{Z_E},~~ {\rm where}~~ Z_E\equiv\Tr_E[\exp(-\beta H_E)],
\label{ZSR2}
\end{align}
namely, the partition function of the Brownian motion cannot be naively expressed as the partition function of the total system divided by the partition function $Z_E$ of the reservoir alone. This is very different from the conventional assumption made in many previous works \cite{PH06,PH08,PH09,PH20,nega4,nega5,HC2,Grabert1,MF1,MF2,MF3,MF4,MF5,MF6,Seifert16,Jarzynski17} where one believes that the reservoir is large enough so that its state changes can be ignored. However, as we will show in the next section, it is this widely accepted but incorrect assumption that generates the inconsistent and controversial results in the studies of quantum thermodynamics for the quantum Brownian motion, such as the occurrence of negative values of heat capacity \cite{PH06,PH08,PH09,nega4,nega5}.

Furthermore, the reduced density matrix of Eq.~(\ref{rho_S}) can be expressed as the standard form of the Gibbs state in terms of the   reduced Hamiltonian $H_S^r$
\begin{align}
\label{rho_Gibbs}
&\rho_S  =\dfrac{1}{Z^r_S}\exp\left[-\dfrac{1}{k_B T}H_S^r\right],
\end{align}
where the reduced Hamiltonian $H_S^r$ can be written as
\begin{align}
&H_S^r=\dfrac{\hbar}{2}[\omega_S^r(a^\dagger a+aa^\dagger)+\Delta_S^r aa+\Delta_S^{r*} a^\dagger a^\dagger]  .\label{renH}
\end{align}
By comparing it with Eq.~(\ref{rho_S}),  the coefficients in the reduced Hamiltonian are simply given by 
\begin{align}
\label{rq}
\hbar\omega_S^r=-2 \eta k_BT ,\ \hbar\Delta^{r}_S=-2\delta k_BT .  
\end{align} 
These two coefficients $\omega_S^r$ and $\Delta_S^r$ are indeed the renormalized frequency and new-induced pairing strength to the Brownian motion 
that are modified and generated by the linear coupling between system and the reservoir. This reduced Hamiltonian is the same as the one 
we obtained from a recent rigorous derivation of the exact master equation for generalized quantum Brownian motion \cite{YW2022}. While, the
paring (squeezing) term in Eq.~(\ref{renH}) is misinterpreted as a part of the dissipation in the the Hu-Paz-Zhang master equation \cite{HPZ1992}, as we have pointed out in our recent work \cite{YW2022}.

Apparently, the reduced Hamiltonian (\ref{renH}) seems to depend explicitly on the temperature through Eq.~(\ref{rq}).
But actually it does not! These two coefficients $\omega_S^r$ and $\Delta_S^r$ are independent of the temperature. 
To see it clearly, we take a Bogoliubov transformation to diagonalize the renormalized 
Hamiltonian $H_S^r$,  
\begin{align}
& \begin{pmatrix}
c\\
c^\dagger
\end{pmatrix}=\begin{pmatrix}
u&v\\
v^*&u^*
\end{pmatrix}\begin{pmatrix}
a\\
a^\dagger
\end{pmatrix}, 
\end{align}
so that
\begin{align}
& H_S^r=\frac{\hbar}{2}\bar{\omega}_S(c^\dagger c+c c^\dagger), \label{dia_H}
\end{align}
where the eigen-frequency of the Bogoliubov quasi-particle is found to be
\begin{align}
\bar{\omega}_S=\sqrt{\omega_S^{r2}-|\Delta_S^r|^2}, \label{renf}
\end{align} 
and the explicit Bogoliubov transformation is given by
\begin{subequations}
\begin{align}
\label{bogo}
&u=\frac{\Delta_S^{r*}}{\sqrt{2(\omega_S^r\bar{\omega}_S-\omega_S^{r2}+|\Delta|^2)}}\\
&v=\frac{\Delta_S^r}{\sqrt{2(\omega_S^r\bar{\omega}_S+\omega_S^{r2}-|\Delta|^2)}}.
\end{align}
\end{subequations}
Now, the relation between occupation $n$ and squeezing $s$ with the renormalized frequency $\omega_S^r$ and pairing strength $\Delta_S^r$ can be obtained from the relation Eq.~(\ref{rq}) and the eigen-frequency of the quasi-particle Eq.~(\ref{renf}),
\begin{align}
\ln\left[\dfrac{\sqrt{\left\vert(n+\frac{1}{2})^2-|s|^2\right\vert}\!-\!\frac{1}{2}}{\sqrt{\left\vert(n+\frac{1}{2})^2-|s|^2\right\vert}\!+\!\frac{1}{2}}\right]=-\dfrac{\hbar\bar{\omega}_S}{k_BT}.
\end{align} 
and 
\begin{subequations}
\label{BE_distribution}
\begin{align}
&n+\frac{1}{2}=\dfrac{\omega^r_S}{\bar{\omega}_S}\left(\dfrac{1}{\exp\left[\frac{\hbar\bar{\omega}_S}{k_BT}\right]-1}+\dfrac{1}{2}\right) ,\\
&s=-\dfrac{\Delta^{r*}_S}{\bar{\omega}_S}\left(\dfrac{1}{\exp\left[\frac{\hbar\bar{\omega}_S}{k_B T}\right]-1}+\dfrac{1}{2}\right) .
\end{align}
\end{subequations}
It gives the particle number occupation $n$ and the squeezing parameter $s$ in the thermodynamic equilibrium state of Eq.~(\ref{rho_Gibbs}) for the system described by the reduced Hamiltonian of Eq.~(\ref{dia_H}). The temperature dependence is all contained in the generalized distribution function incorporating the 
co-existence of the occupation $n$ and the squeezing $s$. The temperature independence of the renormalized frequency and new-induced pairing 
strength, $\omega_S^r$ and $\Delta_S^r$, will also be numerically examined in Fig.~\ref{Fig2}(b) in the next section.

In the Bogoliubov quasi-particle picture, the reduced density matrix Eq.~(\ref{rho_Gibbs}) is a standard thermal state for the effective harmonic oscillator of Eq.~(\ref{dia_H}). The particle number occupation of the Bogoliubov quasi-particle is given by
\begin{align}
n_c&\equiv\langle c^\dagger c\rangle=|u|^2n+|v|^2(n+1)+uv^*s+vu^*s^*\notag\\
&=\sqrt{\left(n+\tfrac{1}{2}\right)^2-|s|^2}-\frac{1}{2}.
\end{align}
Combining this result with Eq.~(\ref{BE_distribution}) and Eq.~(\ref{bogo}), the Bogoliubov quasi-particle number occupation 
in terms of its eigen-frequency $\bar{\omega}_S$ is given by
\begin{align}
n_c=\dfrac{1}{\exp\left[\cfrac{\hbar\bar{\omega}_S}{k_BT}\right]-1}.
\end{align}
That is, only the Bogoliubov quasi-particle can obey the standard Bose-Einstein distribution, whereas the original occupation $n= \Tr[a^\dag a\rho_S]$ does not. The original occupation in the Brownian motion must be described by the extended Bose-Einstein distribution, as shown by Eq.~(\ref{BE_distribution}).

The above results show that after taking the partial trace over all the reservoir states, the system-reservoir interaction not only 
modifies the frequency of the Brownian motion, but also generates the pairing terms for the reduced Hamiltonian of the Brownian motion, as shown in Eq.~(\ref{rho_Gibbs}). 
This pairing term does not exist in the original system Hamiltonian $H_\textsc{S}$ in Eq.~(\ref{H2}).
To make its physical picture clearer, let us rewrite the reduced Hamiltonian Eq.~(\ref{renH}) in terms of the position and momentum operators. The result is
\begin{align}
\label{H_XP}
H_S^r=&\dfrac{1}{2M'}P^2+\dfrac{M'}{2}\left(\omega_S^{r2}-\textrm{Re}[\Delta_S^r]^2\right) X^2\notag\\
&+\dfrac{1}{2}\textrm{Im}[\Delta_S^r](XP+PX),
\end{align}
where $M'=M\omega_S/(\omega_S^r-\textrm{Re}[\Delta_S^r])$. The renormalized pairing
strength $\Delta_S^r$ is related to the squeezing $s$ which is a complex number, and thus the imaginary part of $\Delta_S^r$ cannot be zero.
As a consequence, Eq.~(\ref{H_XP}) shows that the renormalization due to the system-reservoir coupling 
not only shifts the frequency and changes the effective mass of the Brownian motion, but also induces a momentum-dependent 
potential.
The squeezing pairing interaction has not be recognized in the previous studies \cite{Leggett1983,HPZ1992}. Also, the momentum-dependent 
potential is omitted in all the recent investigations to the strong-coupling quantum thermodynamics \cite{PH20,HC2,Grabert1,MF1,MF2,MF3,MF4,MF5,MF6}.
Our result shows further that after traced over all the reservoir states, the state of the system can be expressed as the standard Gibbs state with the reduced Hamiltonian Eq.~(\ref{H_XP}), 
where the original Brownian particle is no longer a harmonic oscillator because it is additionally subjected to a squeezing pairing interaction induced by the system-reservoir coupling. 
 
Furthermore, in terms of the Bogoliubov quasi-particle, the reduced Hamiltonian Eq.~(\ref{renH}) can be represented as an effective 
harmonic oscillator  in a new position and momentum coordinates ($\bar{X}$ and $\bar{P}$), 
\begin{align}
H_S^r=\frac{1}{2M'}\bar{P}^2+\frac{1}{2}M'\bar{\omega}_S^2\bar{X}^2.
\end{align}
The coordinate transformation relation between the original Brownian particle $(X,P)$ and the Bogoliubov quasi-particle
$(\bar{X},\bar{P})$ is given by
\begin{align}
&\begin{pmatrix}
\bar{X}\\
\bar{P}
\end{pmatrix}=\sqrt{\dfrac{M'\bar{\omega}_S}{M\omega_S}}\notag\\
&\times\!\begin{pmatrix}
(\textrm{Re}[u]\!+\!\textrm{Re}[v])\dfrac{M\omega_S}{M'\bar{\omega}_S} &
(\textrm{Im}[v]\!-\!\textrm{Im}[u])\dfrac{1}{M'\bar{\omega}_S}\\
(\textrm{Im}[u]\!+\!\textrm{Im}[v])M\omega_S & (\textrm{Re}[u]\!-\!\textrm{Re}[v])
\end{pmatrix}\!\begin{pmatrix}
X\\ P
\end{pmatrix}
.\end{align}
Through the diagonalization, the system can be viewed as a new effective harmonic oscillator in terms of Bogoliubov quasi-particle only.
The transformation relation between the original and new coordinate is determined by the renormalized frequency and the renormalized pairing strength in the reduced Hamiltonian and is independent of temperature. This result is different from the temperature-dependent Hamiltonian of mean force \cite{PH20,MF1,MF2,MF3,MF4,MF5,MF6} and the early work by Grabert {\it et al.}~\cite{Grabert1}, where they proposed that the reduced density matrix of the Brownian motion is given by a simple harmonic oscillator in the original coordinate with the temperature-dependent effective mass and frequency, which should be incorrect, as we show above. 

As a final check, if we let the squeezing parameter $s=0$ (no pairing in the original Hamiltonian Eq.~(\ref{H2})), then Eq.~(\ref{rho_S}) is reduced to
\begin{align}
\rho_S  |_{s=0} &=\dfrac{1}{1+n}\exp\left\{\ln\left[\frac{n}{1+n}\right]a^\dagger a\right\} \notag \\
&= \dfrac{1}{Z_S} \exp\left\{ - \frac{1}{k_B T}H_{\rm SH} \right\}.
\end{align}
where
\begin{subequations}
\begin{align}
&n=  \Tr[a^\dag a \rho_S] = 1/(e^{\hbar\omega_S^r/k_BT}-1) , \\
&H_{\rm SH}= \hbar \omega^r_S a^\dag a .
\end{align}
\end{subequations}
This reproduces the solution in our previous work \cite{WM1}, in which the coupling Hamiltonian $H_{SB}$ does not contain the paring interaction $\hbar V_k(a^\dagger b_k^\dagger+b_k a)$. 

In conclusion, with the linear couplings between the system and the reservoir \cite{Feynman1963,Leggett1983}, the reduced density matrix of the Brownian motion at equilibrium is given by a Gibbs state with the reduced Hamiltonian that is temperature independent but it must contain a squeezing pairing interaction, as shown by Eq.~(\ref{H_XP}). A momentum-dependent potential which is a part of a pairing interaction is naturally induced by the position-position linear coupling between the system and the reservoir. This result is also obtained from the exact master equation of the generalized quantum Brownian motion we recently derived \cite{YW2022}. 
Note that both the well-known Caldeira-Leggett master equation \cite{Leggett1983} and Hu-Paz-Zhang master equation \cite{HPZ1992}  for quantum Brownian motion did not correctly obtain the squeezing pairing interaction in their reduced Hamiltonian.  This has been justified explicitly in our previous derivation of the exact master equation, see Sec.~IV of  Ref.~\cite{YW2022}. We will further demonstrate the effect of missing the squeezing pairing interaction or the momentum-dependent potential in calculating quantum thermodynamic quantities in the next Section. 

\section{The strong-coupling quantum thermodynamics}
\label{sec3}
In the strong system-reservoir coupling regime, the definitions of thermodynamic quantities are ambiguous. Controversial results have been arisen from various incomplete and inconsistent calculations of the system-environment coupling effects. 
In the previous section, we solve exactly the reduced density matrix that characterizes fully quantum mechanically all the micro-states of the Brownian motion. Thus, various thermodynamic quantities of the Brownian motion can be unambiguously defined. In this section, we shall study the quantum thermodynamics of the Brownian motion based on this exact solution. In particular, we will calculate the particle number occupation, the internal energy, and the heat capacity of the Brownian motion incorporating the squeezing effect and analyze the consistency of the results. Also, we will make comparison with the previous results obtained in other studies.

\subsection{The particle number occupation and squeezing from week to strong couplings}

Note that the particle number occupation in quantum Brownian motion containing pairing interaction does not simply obey the standard 
Bose-Einstein distribution. This is because both the energy frequency $\omega^r_S$ and the pairing strength $\Delta^r_S$ contributes 
to the occupation in the quantum Brownian motion, as shown by Eq.~(\ref{BE_distribution}). 
In Eq.~(\ref{NS}), we have given the solutions of the occupation and squeezing from the total equilibrium state of the system 
and its reservoir for arbitrary reservoir spectrum with arbitrary system-reservoir coupling.
For a practical numerical calculation and without loss of generality, we consider here a Lorentz-Drude spectral density
\begin{align}
\label{J}
J(\omega)\equiv2\pi\sum\limits_k\left\vert V_k\right\vert^2\delta(\omega-\omega_k)=\dfrac{\Gamma}{\omega_S}\dfrac{\omega\omega_D^2}{\omega^2+\omega_D^2},
\end{align}
which has been used in many previous studies of quantum Brownian motion \cite{PH06,PH08,PH09,nega4,nega5,HC2,Grabert1,MF2,Grabert2,drude1,drude2,drude3}. The parameter $\omega_D$ is a cut-off frequency of the reservoir spectrum and $\Gamma$ represents the coupling strength between the system and reservoir. Note that usually numerical results are cut-off frequency dependent, and a counter term is often introduced to avoid the divergence when one takes the cut-off frequency go to infinite \cite{Leggett1983}. However, as we have pointed out \cite{YW2022}, the cut-off frequency introduced in the spectral density defines the effective spectral width of the energy exchange between the Brownian particle and the reservoir. One cannot let the cut-off frequency go to infinite because the energy transfer between the Brownian particles and the high-frequency modes of the reservoir must be drastically reduced. Therefore, physically the counter-term does not need to be considered. 

\begin{figure}
\includegraphics[scale=0.48]{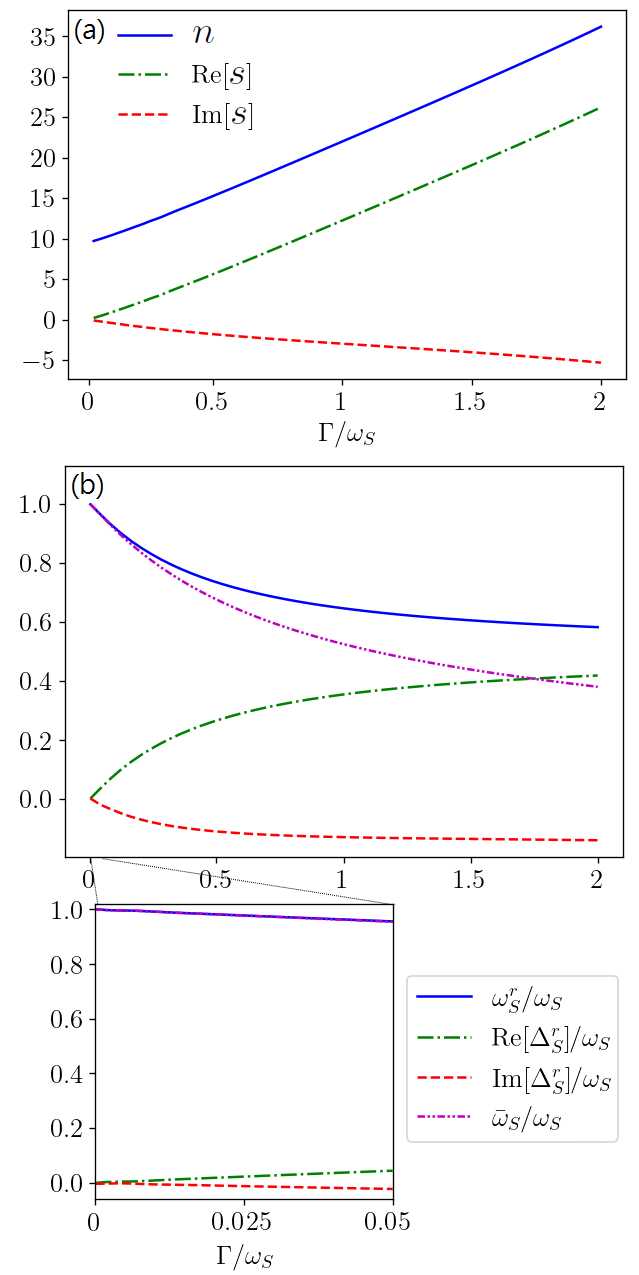}
\caption{(Colour online) (a) The particle number occupation $n$ and squeezing parameter $s$, and (b) the renormalized frequency $\omega^r_S$, the renormalized pairing strength $\Delta^r_S$ and the renormalized eigen-frequency $\bar{\omega}_S$ as functions of the coupling strength $\Gamma/\omega_S$ and its partial enlarged plot at the bottom that focuses on the weak coupling regime $\Gamma/\omega_S < 0.05 $.  Here $n$, $s$ are dimensionless quantities while $\omega^r_S$, $\Delta^r_S$ and $\bar{\omega}_S$ are scaled with unit $\omega_S$. The equilibrium temperature takes $k_BT=10 \hbar \omega_S$ and the cutoff frequency is set as $\omega_D=20 \omega_S$.\label{Fig1}}
\end{figure}

We calculate the particle number occupation $n$ and the squeezing parameter $s$ in the Brownian motion from Eq.~(\ref{NS}) with the system-reservoir coupling characterized by Eq.~(\ref{J}). The numerical results 
of those as functions of the coupling strength $\Gamma/\omega_S$ are plotted in Fig.~\ref{Fig1}(a). 
It shows that a finite system-reservoir coupling can make a non-trivial squeezing effect to the Brownian motion. Only in the very weak-coupling limit $\Gamma/\omega_S \rightarrow 0$, does the value of the squeezing parameter approach to zero, rendering it negligible in the comparison with the particle number occupation. 
In Fig.~\ref{Fig1}(b), we plot the renormalized system frequency $\omega_S^r$, the pairing strength $\Delta_S^r$, and the renormalized eigen-frequency $\bar{\omega}_S=\sqrt{\omega_S^{r2}-|\Delta_S^r|^2}$ as functions of the coupling strength $\Gamma/\omega_S$. It shows that increasing the system-reservoir coupling reduces the renormalized frequency and generates the pairing energy. One can also find that when the system-reservoir coupling is very weak, the renormalized pairing strength  $\Delta_S^r$ becomes significantly less than the renormalized system frequency $\omega_S^r$. It can be seen more clearly in its partial enlarged plot in the bottom of Fig.~\ref{Fig1}(b), where we focus on the weak-coupling regime ($\Gamma/\omega_S < 0.05 $). It shows that the renormalized eigen-frequency $\bar{\omega}_S$ is almost the same with the renormalized frequency $\omega_S^r$ in the very weak coupling regime, while the effect of the pairing strength $\Delta_S^r$ is so weak that it can be neglected in the weak coupling. Thus, only in the weak-coupling regime, the reduced Hamiltonian shown in Eq.~(\ref{renH}) or Eq.~(\ref{H_XP}) can be approximated to a harmonic oscillator with the renormalized frequency $\omega_S^r$ alone. 
In the strong-coupling regime, the pairing energy becomes comparable to that of the renormalized frequency  $\omega_S^r$. As a result, it is necessary to take into account the contribution of the pairing effect in Eq.~(\ref{renH}) or the momentum-dependent potential shown in Eq.~(\ref{H_XP}) for the calculations of all thermodynamic quantities in the strong coupling regime. 

\begin{figure}
\includegraphics[scale=0.4]{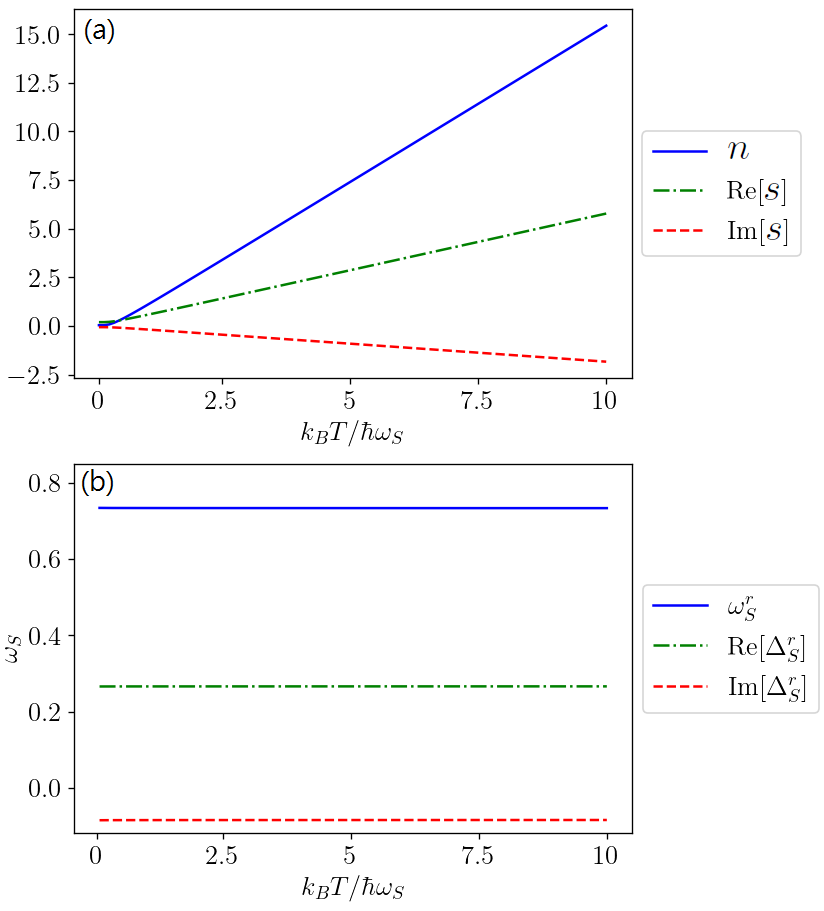}
\caption{(Colour online) (a) The particle number occupation and particle squeezing and (b) the renormalized frequency and the renormalized pairing strength as functions of temperature with the coupling strength $\Gamma/\omega_S=0.5 $. The latter two quantities are independent of temperature, as shown in Fig.~\ref{Fig2}(b), so does the reduced Hamiltonian Eq.~(\ref{renH}) or Eq.~(\ref{H_XP}).\label{Fig2}}
\end{figure}

In Fig.~\ref{Fig2}, we also present those physical quantities as functions of the dimensionless temperature $k_BT/\hbar \omega_S$. Figure~\ref{Fig2}(a) shows the monotonic variations of both the particle number occupation $n$ and squeezing parameter $s$ as the temperature increasing. Contrastingly, Fig.~\ref{Fig2}(b) reveals that both the renormalized frequency $\omega_S^r$ and pairing strength $\Delta_S^r$ remain unchanged with respect to the temperature change, although they are calculated based on the temperature-dependent particle number occupation $n$ and squeezing parameter $s$ through Eq.~(\ref{12}) and Eq.~(\ref{rq}). This justifies our conclusion made in Sec.~II that the  reduced Hamiltonian $H_S^r$ derived from the exact reduced density matrix of Eq.~(\ref{rho_Gibbs}) is temperature independent, in contrast to the Hamiltonian of mean force introduced in previous works \cite{PH20,MF1,MF2,MF3,MF4,MF5,MF6} which is considered or  assumed to be temperature-dependent.

\subsection{The renormalized internal energy and heat capacity}
\subsubsection{The calculations with the exact reduced density matrix and the exact reduced partition function}
The definition of the system internal energy beyond weak coupling has been debated for many years. In the literature, different definitions of the internal energy give inconsistent results \cite{PH06,PH08,PH09}. We find that the inconsistency comes from the lack of the way to take into account all the renormalization effects induced by the system-reservoir coupling. With the exact solution of the reduced density matrix $\rho_S$ given by Eq.~(\ref{rho_Gibbs}) [or Eq.~(\ref{rho_S})] and the reduced Hamiltonian of Eq.~(\ref{renH}), which encapsulates all the renormalization effects from the system-reservoir coupling, the internal energy of the Brownian motion can be defined unambiguously and self-consistently. The direct definition of the internal energy is given by
\begin{align}
\label{UHa}
U_H\equiv&\Tr_S[H_S^r\rho_S]\notag\\
=&\dfrac{1}{2M'}\langle P^2\rangle+\dfrac{M'}{2}\left(\omega_S^{r2}-\textrm{Re}[\Delta_S^r]^2\right) \langle X^2\rangle\notag\\
&+\dfrac{1}{2}\textrm{Im}[\Delta_S^r]\langle XP+PX\rangle\notag\\
=&\dfrac{\hbar}{2}\big[\omega_S^r(2n+1)+2\textrm{Re}[\Delta_S^rs]\big]
\end{align}
With the aid of Eq.~(\ref{BE_distribution}), the internal energy can be reduced to
\begin{align}
\label{UH}
U_H=\hbar\bar{\omega}_S\left(\dfrac{1}{\exp\left[\frac{\hbar\bar{\omega}_S}{k_BT}\right]-1}+\frac{1}{2}\right).
\end{align}
It shows that the internal energy can be expressed in terms of the renormalized eigen-frequency $\bar{\omega}_S$ of the Bogoliubov quasi-particle. The Bogoliubov quasi-particle is 
a mixture of the original harmonic oscillator with all of particles in the reservoir, as shown by Eq.~(\ref{renf}).

Alternatively, one can also define the internal energy from the fully reduced partition function $Z_S^r$ of the  quantum Brownian motion, given by Eq.~(\ref{eZSR}) which can be expressed further as
\begin{align}
Z_S^r&=\sqrt{n^2+n-|s|^2}=\dfrac{1}{2}\csch\left(\dfrac{\hbar\bar{\omega}_S}{2k_BT}\right).
\end{align}
Then using the well-known definition of the  internal energy through the partition function, we obtain
\begin{align}
U_Z&\equiv-\dfrac{\partial}{\partial\beta}\ln(Z_S^r)=\dfrac{1}{2}\hbar\bar{\omega}_S\coth\left(\dfrac{\hbar\bar{\omega}_S}{2k_BT}\right)\notag\\
&=\hbar\bar{\omega}_S\left(\dfrac{1}{\exp\left[\frac{\hbar\bar{\omega}_S}{k_BT}\right]-1}+\frac{1}{2}\right).
\label{UZ}
\end{align}
It shows that the internal energy $U_Z$ defined through the fully reduced partition function is exactly the same as the internal energy $U_H$ defined by the reduced Hamiltonian with the exact solution of the reduced density matrix, as shown in Eq.~(\ref{UH}). In other words, the two different definitions of the internal energy agree with each other, unlike in the previous works \cite{PH06,PH08,PH09} where they used an incomplete-reduced Hamiltonian and an incorrect partition function that result in  inconsistent solutions to the internal energy. 

\begin{figure}
\includegraphics[scale=0.48]{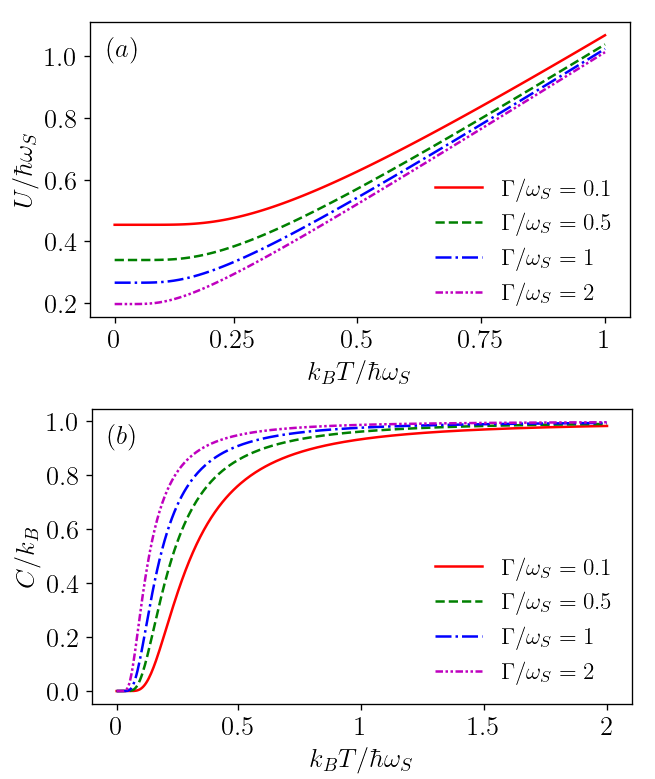}
\caption{(Colour online) (a) The internal energy and (b) the heat capacity as a function of the temperature with several different coupling strength $\Gamma/\omega_S$, calculated from both the fully reduced Hamiltonian with the exact reduced density matrix Eq.~(\ref{UHa}) and the fully reduced partition function of Eq.~(\ref{UZ}), both give the same result. \label{Fig3}}
\end{figure}

With the self-consistent result of internal energy presented by Eq.~(\ref{UH}) and Eq.~(\ref{UZ}), we can obtain the unique heat capacity for quantum Brownian motion
\begin{align}
\label{C}
C\equiv\dfrac{\partial U}{\partial T}=\left[\dfrac{\hbar\bar{\omega}_S}{2k_BT}\csch\left(\dfrac{\hbar\bar{\omega}_S}{2k_BT}\right)\right]^2.
\end{align}
It is also consistent with the results of the Einstein model for a single-mode quantum Brownian motion, except that the harmonic oscillator frequency is now given by the renormalized eigen-frequency of the Bogoliubov quasi-particle, which takes into account all the renormalization effects of the system-reservoir coupling on the Brownian motion, including the frequency shift and the induced pairing (squeezing) interaction. 
The result of Eq.~(\ref{C}) is a monotonically increasing function of the ratio $k_BT/\hbar \bar{\omega}_S$, and obviously in no case does it lead to a negative heat capacity obtained in previous studies \cite{PH06,PH08,PH09,nega4,nega5}. At low temperatures, it decreases to zero as $C\simeq(\frac{\hbar\bar{\omega}_S}{2k_BT})^2\exp(-\frac{\hbar\bar{\omega}_S}{k_BT})$, while it approaches to the Boltzmann constant $k_B$ at high temperatures, as one expected. 

In Fig.~\ref{Fig3}, we present the numerical results of the internal energy and the heat capacity as a function of the temperature with several 
different coupling strength.  The internal energy as a function of temperature with several different coupling strength $\Gamma/\omega_S$ is shown in Fig.~\ref{Fig3}(a). At very low temperature, the different coupling strengths make the internal energy significantly different.
On the other hand, for the reservoir with the Drude-Lorentz spectral density, the renormalized eigen-frequency $\bar{\omega}_S=\sqrt{\omega_S^{r2}-|\Delta_S^r|^2}$ always decrease as the coupling strength $\Gamma/\omega_S$ increases, as shown in Fig.~\ref{Fig1}(b), 
Thus, the heat capacity will increase monotonically with the coupling strength increasing, as shown in Fig.~\ref{Fig3}(b) where the heat capacity is presented as a function of the temperature for  
different coupling strengths.
At high temperature, the internal energy and the heat capacity of Brownian motion approach to the classical limit, following the equipartition theorem, as one expected. In other words, a one-dimensional harmonic oscillator has the average energy $1\ k_B T$, and therefore its heat 
capacity is $1\ k_B$, as shown  in Fig.~\ref{Fig3}(a) and Fig.~\ref{Fig3}(b) for high temperature. Thus the effects of the system-reservoir coupling become negligible in high temperature, reproducing the classical thermodynamic solution for Brownian motion.

\subsubsection{Comparing our results with previous works}

\begin{figure}
\includegraphics[scale=0.48]{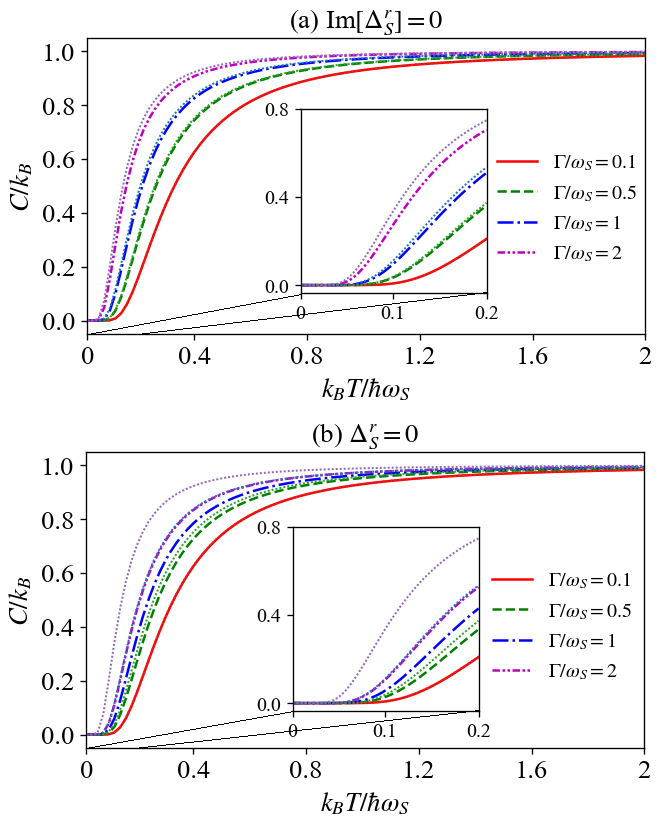}
\caption{(Colour online) The heat capacity as a function of temperature with different coupling strength $\Gamma/\omega_S$. (a) The imaginary part of $\Delta_S^r$ is zero, namely the momentum-dependent potential in the reduced Hamiltonian is ignored, and (b) both the imaginary part and the real part of $\Delta_S^r$ are set to be zero, namely the squeezing effect is totally ignored. The light dotted lines are the  results (based on our fully renormalized theory with both $\textrm{Im}[\Delta_S^r]\neq0$ and $\textrm{Re}[\Delta_S^r]\neq0$) as given in Fig.~\ref{Fig3}(b), serving as a basis for comparison. \label{Fig4}}
\end{figure}

In the preceding discussion, in order to unambiguously determine thermodynamic quantities in the framework of quantum thermodynamics, 
we have elucidated the necessity of taking into account all the renormalization effects on the Brownian motion Hamiltonian as well as its partition function, induced by the system-reservoir coupling. Now, we would like to compare our results with the results obtained by others based on an incomplete-reduced Hamiltonian and an incorrect partition function of the Brownian motion.
Here the term ''incomplete-reduced Hamiltonian" refers to the ignorance of the the system-reservoir coupling-induced momentum-dependent potential (or the pairing effect) in the previous works \cite{HC2,Grabert1,MF2,PH06,PH08}. 
The incorrect partition function corresponds to the thought which has been widely used in the literature that the reservoir is large enough so its states should remain unchanged. Thus, the partition function of the system was naively defined as the partition function of the total system divided by that of the reservoir: $Z_S = Z_{\rm tot}/Z_E$, where $Z_E\equiv \Tr_E[\exp(-\beta H_E)]$, as used in many previous works \cite{PH06,PH08,PH09,PH20,nega4,nega5,HC2,MF1,MF2,MF3,MF4,MF5,MF6,Seifert16,Jarzynski17}. However, such widely used definition of the  partition function for the study of quantum thermodynamics of any open quantum systems is indeed not always correct. Our derivation in Sec.~\ref{sec2} shows that the reservoir and the system will mutually influence each other, which non-negligibly change both the system and the reservoir states.

In Fig.~\ref{Fig4}(a), we plot the heat capacity for different coupling strengths calculated by the incomplete-reduced Hamiltonian which ignores the momentum-dependent potential, as considered in Ref.~\cite{HC2,Grabert1,MF2}. In Fig.~\ref{Fig4}(b), we also plot the heat capacity by enforcing both the real and imaginary parts of the renormalized pairing strength $\Delta_S^r$ to be zero in Eq.~(\ref{UHa}) so that not only the momentum-dependent potential energy is omitted, but also the oscillation frequency of the Brownian motion influenced by the thermal reservoir is not renormalized, as considered in Ref.~\cite{PH06,PH08}.  We make a comparison with the results calculated from our theory, i.e., the results of Fig.~\ref{Fig3} and plot them by the light dotted lines in Fig.~\ref{Fig4}(a)-(b). 
In Fig.~\ref{Fig4}(a), the differences are primarily manifested in the strong coupling regime (e.g. $\Gamma/\omega_S > 2 $), where the renormalized pairing strength $\Delta_S^r$ become important and its imaginary part is non-negligible compared to the renormalized eigen-frequency $\bar{\omega}_S$, as shown in Fig.~\ref{Fig1}(b). Note that the heat capacity given by Eq.~(\ref{C}) is totally determined by the ratio of the renormalized eigen-frequency to temperature $\Bar{\omega}_S/T$. Therefore, the difference arisen from the divergence between incomplete renormalizated frequency and $\Bar{\omega}_S$ is more pronounced in the low temperature regime. This is because in the high temperature regime, the pairing dynamics is not important. The differences in the low temperature regime become significant, as shown in Fig.~\ref{Fig4}(b), where the pairing effects plays an important role but they are completely ignored in the incomplete-reduced Hamiltonian \cite{PH06,PH08}. Only in the very weak coupling (e.g. $\Gamma/\omega_S=0.1 $), where the renormalized eigen-frequency $\bar{\omega}_S$ can be approximated by the renormalized frequency $\omega_S^r$ [as shown in Fig.~\ref{Fig1}(b)], where the heat capacity calculated with the incomplete reduced Hamiltonian gives almost the same value as we obtained from our theory [see the red line in Fig.~\ref{Fig4} ].
These results indicate that a correct reduced Hamiltonian is significantly important to define quantum thermodynamical quantities in the strong coupling. 


\begin{figure}
\includegraphics[scale=0.48]{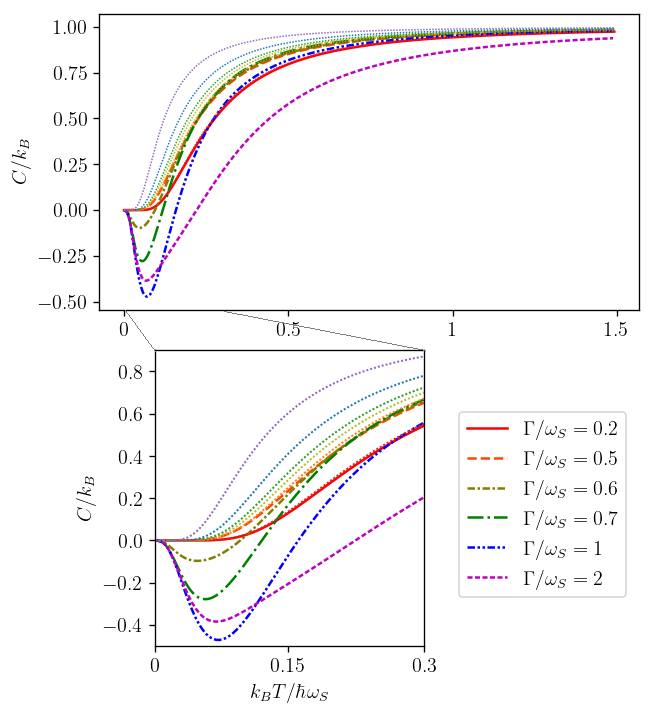}
\caption{(Colour online) The heat capacity as a function of temperature with different coupling strength $\Gamma/\omega_S$, calculated based on the  incorrect partition function $Z_S=Z_{tot}/Z_E$. The light dotted-lines are the results calculated based on the  reduced partition function $Z_S^r$ given in Fig.~\ref{Fig3}(b), serving as a basis for comparison here. \label{Fig5}}
\end{figure}

On the other hand, we also numerically demonstrate why it gives the negative heat capacity in the previous investigations of quantum Brownian motion
\cite{PH06,PH08,PH09,nega4,nega5}. We find that the negative heat capacity comes from the naive assumption of the reduced partition function $Z_S \equiv Z_{\rm{tot}}/Z_E$. This naive assumption has been used widely in the literature because it is commonly thought of the reservoir being larger enough so that 
the reservoir states remain unchanged even if the system strongly couples the reservoir. As we have shown in this work, this assumption is  
incorrect in the strong coupling. The exactly reduced partition function $Z_S^r$ for the quantum Brownian motion is given by Eq.~(\ref{ZSR}) which cannot be written as 
 $Z^r_S = Z_{\rm{tot}}/Z_E$. 
Recall the relation between the internal energy and partition function $U=-\frac{\partial}{\partial\beta}\ln(Z)$ shown in the first line of Eq.~(\ref{UZ}), if one uses the partition function $Z_S=Z_{\rm{tot}}/Z_E$, the corresponding internal energy is given by
\begin{align}
    U &=-\frac{\partial}{\partial\beta}[\ln (Z_{\rm{tot}})-\ln (Z_E)]\notag\\
    &=-\frac{\partial}{\partial\beta}\left[\ln\left(\Tr_{\rm{tot}} e^{-\beta H_{\rm{tot}}}\right)-\ln\left(\Tr_{\rm{E}} e^{-\beta H_E}\right)\right].
    \label{UZP}
\end{align}
Then the heat capacity $C\equiv\dfrac{\partial U}{\partial T}$ based on this incorrect internal energy will give 
negative values in the strong coupling, as have been numerically shown in the early studies \cite{PH06,PH08,PH09,nega4,nega5}, and as we also plot them in Fig.~\ref{Fig5}. It becomes more obvious in the partially enlarged plot in the low temperature regime in Fig.~\ref{Fig5}, where one can clearly see the appearance of the negative heat capacity for strong couplings $\Gamma/\omega_S \geq 0.6$. 
In Fig.~\ref{Fig5}, we also plot the heat capacity based on the fully reduced partition function $Z^r_S$ of  Eq.~(\ref{ZSR}) with the light
dotted lines [also see Fig.~\ref{Fig3}(b)] for a comparison. It shows that the negative heat capacity of the quantum Brownian motion obtained in the previous studies is a consequence of the
use of an incorrect partition function. Thus, despite the reservoir is typically and significantly larger than the system, its states are still changed due to the strong entanglement between the system and the reservoir in the strong coupling. It is necessary to find the renormalization of the reduced partition function and the reduced Hamiltonian of the system in a correct method. Only in the weak coupling regime, the heat capacities calculated based on $Z_S$ and $Z_S^r$ are nearly identical. Therefore, the claimed anomalous phenomenon of negative heat capacity under strong coupling in the literature is, in fact, a misinterpretation resulting from the incorrect calculation of the partition function.

\section{Conclusions and Discussions}
\label{concl}
In conclusion, we apply our nonperturbative renormalization theory to the quantum Brownian motion. We traced over exactly all the reservoir states through the Gaussian integral approach in coherent state representation. Based on the result obtained from these rigorous derivations, we find that the widely used definition of the reduced partition function $Z_S\equiv Z_{\rm{tot}}/Z_E$ for open systems in the literature \cite{PH06,PH08,PH09,PH20,nega4,nega5,HC2,Grabert1,MF1,MF2,MF3,MF4,MF5,MF6,Seifert16,Jarzynski17} is not valid for strong system-reservoir couplings. Our solution indicates that in the strong coupling, even if the thermal reservoir is significantly larger (containing much more degrees of freedom) than the system, the reservoir states are still nontrivially changed by the system through the system-reservoir coupling. We also show that the approach based on the mean force Hamiltonian \cite{PH20,MF1,MF2,MF3,MF4,MF5,MF6,Seifert16,Jarzynski17}, which has been widely used to describe quantum thermodynamics, is actually also not valid in the strong coupling.
Furthermore, utilizing the faithful representation in group theory, we successfully find the correct operator form of the renormalized system Hamiltonian and derive the exact reduced density matrix for equilibrium quantum Brownian motion. The exact reduced density matrix has the standard Gibbs state in terms of the fully reduced Hamiltonian, given by Eqs.~(\ref{rho_Gibbs}) and (\ref{renH}), and it is indeed a squeezing thermal state. The renormalization effects arisen from the system-reservoir coupling not only shifts the frequency $\omega_S^r$ of the renormalized Brownian motion, but also generates the pairing energies $\Delta_S^r aa+\Delta_S^{r*} a^\dagger a^\dagger$ into the reduced Hamiltonian. The existence of pairing interaction inevitably generates a momentum-dependent potential. 
 
As a consequence of the presence of pairing interaction in the reduced Hamiltonian, the relation between  occupation $n$ and renormalized frequency $\omega_S^r$ of the Brownian motion cannot simply obey the Bose--Einstein distribution function. Based on the exact solutions of reduced density matrix and the fully reduced Hamiltonian, we find an extended distribution function that describes the relation between  occupation $n$ and squeezing $s$ in terms of the renormalized frequency $\omega_S^r$ and the renormalized pairing strength $\Delta_S^r$, as given by Eq.~(\ref{BE_distribution}).  Only after applying a Bogoliubov transformation to diagonalize the  reduced Hamiltonian of the Brownian motion, it allows one to represent the Bogoliubov quasi-particle as an effective harmonic potential in the new position-momentum coordinates. Then the occupation of the Bogoliubov quasi-particle 
obeys the conventional Bose--Einstein distribution with the renormalized eigen-frequency $\bar{\omega}_S$ given by Eq.~(\ref{renf}). We also show that the fully reduced Hamiltonian is temperature independent.

We study further the internal energy and heat capacity based on the exact reduced density matrix and the reduced Hamiltonian. We obtain the consistent internal energy for the Brownian motion from the two different definitions based on the fully reduced Hamiltonian and based on the fully reduced partition function, respectively. The controversial results of the heat capacity used these two different definitions obtained in the previous works \cite{PH06,PH08,PH09} are resolved. We also find that the corresponding heat capacity of the Brownian motion agrees with the results of the Einstein model with the distinction that the oscillation frequency must be replaced by the renormalized eigen-frequency of the Bogoliubov quasi-particle. Moreover, we numerically compare our results with those obtained based on the incomplete-reduced partition function and incomplete-reduced Hamiltonian, as considered in the literature. 
We find that for the incomplete reduced Hamiltonian which only neglects the potential related to momentum (ignoring the imaginary part of the particle squeezing parameter $s$), the calculated heat capacity differs little from our results, with discrepancies appearing only in the very strong coupling. However, when both the real and imaginary parts of the squeezing parameter are ignored, the lack of the paring interaction in the incomplete-reduced Hamiltonian leads to a significant difference in the heat capacity at the low temperature regime, where the pairing effect plays an important role in quantum squeezing  thermodynamics.  More importantly, the issue of the negative heat capacity in the low-temperature regime is also resolved. The negativity of the heat capacity comes from the use of an incorrect reduced partition function in the previous works. Our investigation could potentially provide a foundation for the study of the quantum thermodynamics without being constrained by the ambiguous definitions of thermodynamic quantities in quantum mechanics.

\acknowledgments
This work is supported by National Science and Technology Council of Taiwan, Republic of China, under Contract
No. MOST-111-2811-M-006-014-MY3.

\appendix
\begin{widetext}
\section{Derivation of the exact reduced density matrix using the faithful matrix representation}
\label{appendixA}
In this Appendix, we present the derivation of the exact reduced density matrix of the Brownian motion. We start from the particle number representation of the Hamiltonian for quantum Brownian motion as shown by Eq.~(\ref{H2}). This Hamiltonian is indeed a linear function of the generators of the symplectic Lie group $Sp(2(k_c+1)+2)$ for the mutil-mode bosonic systems (see Sec.~D1 of Ref.~\cite{group}), and $k_c$ is the total number of modes in the reservoir.
Because the total system is assumed to be in equilibrium, the density matrix of the total system (the Brownian particle plus its reservoir) should be able to be represented as 
\begin{align}
\rho_{\rm{tot}}&=\dfrac{1}{Z_{\rm{tot}}}e^{-\beta H_{\rm{tot}}}\notag\\
&=\dfrac{1}{Z_{\rm{tot}}}\exp\left\{-\hbar\beta
\left[\frac{\omega_S}{2}(a^\dag a+aa^\dag)+\sum_k\frac{\omega_k}{2}(b^\dag b+bb^\dag)+\sum_k V_k(a^\dag+a)(b^\dag+b)
+\lambda\right]
\right\},
\label{A1}
\end{align}
where $\lambda=-\dfrac{1}{2}\Big(\omega_S + \sum_k  \omega_k\Big)$. The faithful matrix representation \cite{group} of the algebra of the total system gives the operator in terms of matrix as follows: 
\begin{align}
\frac{\omega_S}{2}(a^\dag a+aa^\dag)+\sum_k\frac{\omega_k}{2}(b^\dag b+bb^\dag)+\sum_k V_k(a^\dag+a)(b^\dag+b)+\lambda
=\begin{pmatrix}
0&0&\bm{0}&\bm{0}&0&0\\
0&\omega_S&\bm{V}_{S\bm{E}}&\bm{\bar{V}}_{S\bm{E}}&0&0\\
\bm{0}&\bm{V}^\dag_{S\bm{E}}&\boldsymbol{\omega_E}&\boldsymbol{0}&\bm{V}^\dag_{S\bm{E}}&\bm{0}\\[+1mm]
\bm{0}&-\bm{\bar{V}}^\dag_{S\bm{E}}&\boldsymbol{0}&-\boldsymbol{\tilde{\omega}_E}&-\bm{\bar{V}}^\dag_{S\bm{E}}&\bm{0}\\
0&0&-\bm{V}_{S\bm{E}}&-\bm{\bar{V}}_{S\bm{E}}&-\omega_S&0\\
\lambda&0&\bm{0}&\bm{0}&0&0
\end{pmatrix},
\end{align}
where 
\begin{align}
&\hspace{-3.5mm}\boldsymbol{\omega_E}=\!\begin{pmatrix}
&\omega_{k_1} & & & \\[-2mm]
& & \omega_{k_2}  &  & \hspace{-2mm}\mbox{\Huge 0}\\
& \hspace{-2mm}\mbox{\Huge 0} & & & \hspace{-2mm}\ddots 
\end{pmatrix} \!, \ 
\boldsymbol{\widetilde{\omega}_E}=\!\begin{pmatrix}
&\ddots & & &\hspace{-2mm}\mbox{\Huge 0} \\
& & &\hspace{-2mm} \omega_{k_2}  &\\[-2mm]
&\hspace{-2mm}\mbox{\Huge 0} & & &\omega_{k_1} 
\end{pmatrix} \!, \
&\hspace{-3.5mm}\bm{V}_{S\bm{E}}=\begin{pmatrix}
V_{k_1} & V_{k_2} & \cdots 
\end{pmatrix},\ \bm{\bar{V}}_{S\bm{E}}=\begin{pmatrix}
\cdots& V_{k_2} & V_{k_1}
\end{pmatrix} .
\end{align}
Then the faithful matrix representation of the density matrix of the total system can be expressed as 
\begin{align}
\label{rho_t1}
\rho_{\rm{tot}}=\begin{pmatrix}
1&\bm{0}&0\\
\bm{0}&\exp\left[\begin{array}{cc}
\bm{D}&\bm{R}\\
-\bm{\hat{R}}&-\bm{\tilde{D}}
\end{array}
\right] &\bm{0}\\
2\ln Z_{\rm{tot}}&\bm{0}&1
\end{pmatrix},\ 
\bm{D}=-\dfrac{\hbar\beta}{2}\begin{pmatrix}
\omega_S&\bm{V}_{S\bm{E}}\\
\bm{V}_{S\bm{E}}^\dag&\boldsymbol{\omega_{E}}
\end{pmatrix},\
\bm{R}=-\dfrac{\hbar\beta}{2}\begin{pmatrix}
\bm{\bar{V}}_{S\bm{E}}&0\\
\bm{0}&\bm{V}_{S\bm{E}}^\dag
\end{pmatrix},
\end{align}
where the "tilde"  indicates reflection in the minor diagonal, and the "hat"  indicates reflection in the major diagonal. 

Now we rearrange the ordering of the exponential operator products of Eq.~(\ref{A1}) as
\begin{align}
\label{rho_t02}
\rho_{\rm{tot}}=&\dfrac{1}{Z_{\rm{tot}}}\exp\left[\dfrac{1}{2}\begin{pmatrix}
a^\dag & \boldsymbol{b^\dag}
\end{pmatrix}\begin{pmatrix}
\Pi_{SS}&\bm{\Pi}_{S\bm{E}}\\
\bm{\Pi}_{S\bm{E}}^*&\bm{\Pi_{EE}}
\end{pmatrix}\begin{pmatrix}
a^\dag\\\boldsymbol{b^\dag}
\end{pmatrix}\right]\notag\\
&\times\exp\left[\dfrac{1}{2}\begin{pmatrix}
a^\dag & \boldsymbol{b^\dag}
\end{pmatrix}\begin{pmatrix}
\Omega^{_{0}}_{SS}&\bm{\Omega^{_{0}}}_{S\bm{E}}\\
\bm{\Omega}_{S\bm{E}}^{_{\bm{0}}*}&\bm{\Omega^{_{0}}_{EE}}
\end{pmatrix}\begin{pmatrix}
a\\\boldsymbol{b}
\end{pmatrix}
+\dfrac{1}{2}\begin{pmatrix}
a & \boldsymbol{b}
\end{pmatrix}\begin{pmatrix}
\Omega^{_{0}}_{SS}&\bm{\Omega}^{_{\bm{0}}*}_{S\bm{E}}\\
\bm{\Omega^{_{0}}}_{S\bm{E}}&\bm{\Omega^{_{0}}}_{\bm{EE}}
\end{pmatrix}\begin{pmatrix}
a^\dag\\\boldsymbol{b^\dag}
\end{pmatrix}\right]\notag\\
&\times\exp\left[\dfrac{1}{2}\begin{pmatrix}
a & \boldsymbol{b}
\end{pmatrix}\begin{pmatrix}
\Pi_{SS}&\bm{\Pi}_{S\bm{E}}^*\\
\bm{\Pi}_{S\bm{E}}&\bm{\Pi_{EE}}
\end{pmatrix}\begin{pmatrix}
a\\\boldsymbol{b}
\end{pmatrix}\right].
\end{align}
Its faithful matrix representation is explicitly given by
\begin{align}
\label{rho_t2}
&\rho_{\rm{tot}}=\begin{pmatrix}
1&\bm{0}&\bm{0}&0\\
\bm{0}&\boldsymbol{\Omega}-\,\boldsymbol{\Pi}\,\boldsymbol{\tilde{\Omega}}^{-1}\boldsymbol{\hat{\Pi}}&
\boldsymbol{\Pi}\,\boldsymbol{\tilde{\Omega}}^{-1}&\bm{0}\\
\bm{0}&-\boldsymbol{\tilde{\Omega}}^{-1}\boldsymbol{\hat{\Pi}}&\boldsymbol{\tilde{\Omega}}^{-1}&\bm{0}\\
2\ln Z_{\rm{tot}}&\bm{0}&\bm{0}&1
\end{pmatrix},\ 
\end{align}
where
\begin{align}
&\boldsymbol{\Pi}=\begin{pmatrix}
\Pi_{SS}&\bm{\Pi}_{S\bm{E}}\\
\bm{\Pi}_{S\bm{E}}^*&\bm{\Pi_{EE}}
\end{pmatrix},\
\boldsymbol{\Omega}=\begin{pmatrix}
\Omega_{SS}&\bm{\Omega}_{S\bm{E}}\\
\bm{\Omega}_{S\bm{E}}^{*}&\bm{\Omega_{EE}}
\end{pmatrix}=\exp\left[\begin{pmatrix}
\Omega^{_{0}}_{SS}&\bm{\Omega^{_{0}}}_{S\bm{E}}\\
\bm{\Omega}_{S\bm{E}}^{_{\bm{0}}*}&\bm{\Omega^{_{0}}_{EE}}
\end{pmatrix}\right].
\end{align}
By comparing Eq.~(\ref{rho_t1}) and Eq.~(\ref{rho_t2}), one can find that
\begin{align}
\bm{\Tilde{\Omega}}=\left\{\begin{pmatrix}
\bm{0}&\bm{I}
\end{pmatrix}\exp\left[\begin{pmatrix}
\bm{D}&\bm{R}\\
-\bm{\hat{R}}&-\bm{\tilde{D}}
\end{pmatrix}
\right]\begin{pmatrix}
\bm{0}\\ \bm{I}
\end{pmatrix}\right\}^{-1},\
\bm{\Pi}=\left\{\begin{pmatrix}
\bm{I}&\bm{0}
\end{pmatrix}\exp\left[\begin{pmatrix}
\bm{D}&\bm{R}\\
-\bm{\hat{R}}&-\bm{\tilde{D}}
\end{pmatrix}
\right]\begin{pmatrix}
\bm{0}\\ \bm{I}
\end{pmatrix}\right\}\bm{\tilde{\Omega}},
\end{align}
as given in Eq.~(\ref{Omega_Pi}). The coherent state representation of the density matrix of the total system can be easily obtained
\begin{align}
\langle z_S,\bm{z_E}|\rho_{\rm{tot}}|z_S^\prime,\bm{z_E}\rangle=&\dfrac{1}{Z_{\rm{tot}}}\exp\left[\dfrac{1}{2}\bm{z^\dag}\bm{\Pi}\bm{z^*}+\dfrac{1}{2}\bm{z^{\prime T}}\bm{\Pi^*}\bm{z'}\right]\notag\\
&\times\left\langle z_S,\bm{z_E}\left|\exp\left[\begin{pmatrix}
a^\dag & \boldsymbol{b^\dag}
\end{pmatrix}\begin{pmatrix}
\Omega^{_{0}}_{SS}&\bm{\Omega^{_{0}}}_{S\bm{E}}\\
\bm{\Omega}_{S\bm{E}}^{_{\bm{0}}*}&\bm{\Omega^{_{0}}_{EE}}
\end{pmatrix}\begin{pmatrix}
a\\\boldsymbol{b}
\end{pmatrix}
+\dfrac{1}{2}\Tr\begin{pmatrix}
\Omega^{_{0}}_{SS}&\bm{\Omega}^{_{\bm{0}}*}_{S\bm{E}}\\
\bm{\Omega^{_{0}}}_{S\bm{E}}&\bm{\Omega^{_{0}}}_{\bm{EE}}
\end{pmatrix}\right]\right|z_S^\prime,\bm{z_E}\right\rangle\notag\\
=&\dfrac{1}{Z_{\rm{tot}}}\exp\left[\dfrac{1}{2}\Tr\ln(\bm{\Omega})\right]
\exp\left[\bm{z^\dag}\bm{\Omega}\bm{z'}+\dfrac{1}{2}\bm{z^\dag}\bm{\Pi}\bm{z^*}+\dfrac{1}{2}\bm{z^{\prime T}}\bm{\Pi^*}\bm{z'}\right]\notag\\
=&\dfrac{\sqrt{\det \bm{\Omega}}}{Z_{\rm tot} }\exp\!\left[\dfrac{1}{2}\begin{pmatrix}
\boldsymbol{z^\dagger} & \boldsymbol{z^{\prime T}}\end{pmatrix}\!\!
\begin{pmatrix}
\boldsymbol{\Omega}  & \boldsymbol{\Pi} \\
\boldsymbol{\Pi^*}  &\boldsymbol{\Omega^*} 
\end{pmatrix}\!\!
\begin{pmatrix}
\boldsymbol{z^\prime}\\
\boldsymbol{z^*}
\end{pmatrix}\!
\right],
\label{A9}
\end{align}
as given by Eq.~(\ref{rho_tot}). 

Taking the partial trace over all the reservoir modes, we obtain the reduced density matrix of the Brownian motion in the coherent state representation
\begin{align}
\label{rhos_coherent}
\langle z_S|\rho_S
|z^\prime_S\rangle=&
\int d\mu(\boldsymbol{z_E})\langle z_S,\boldsymbol{z_E}|
\rho_{\rm{tot} }
|z_S^\prime,\boldsymbol{z_E}\rangle\notag\\
=&\dfrac{\sqrt{\det \bm{\Omega}}}{Z_{\rm{tot}} }\int d\mu(\boldsymbol{z_E})
\exp\left[\dfrac{1}{2}\begin{pmatrix}
\boldsymbol{z^\dagger} & \boldsymbol{z^{\prime T}}\end{pmatrix}
\begin{pmatrix}
\boldsymbol{\Omega}  & \boldsymbol{\Pi} \\
\boldsymbol{\Pi^*}  &\boldsymbol{\Omega^*} 
\end{pmatrix}
\begin{pmatrix}
\boldsymbol{z^\prime}\\
\boldsymbol{z^*}
\end{pmatrix}
\right]
\end{align}
where $d\mu(\boldsymbol{z_E})=\prod\limits_j dz_j^* dz_j e^{-|z_{j}|^2}$. 
The Gaussian integral $\int d\mu(\boldsymbol{\xi})e^{\boldsymbol{\xi^\dagger}\cdot\boldsymbol{\Theta}\cdot\boldsymbol{\xi}+\boldsymbol{\eta^\dagger}\cdot\boldsymbol{\xi}+\boldsymbol{\xi^\dagger}\cdot\boldsymbol{\eta^\prime}}=||\boldsymbol{1}-\boldsymbol{\Theta}||e^{\boldsymbol{\eta^\dagger}\cdot(\boldsymbol{1}-\boldsymbol{\Theta})^{-1}\cdot\boldsymbol{\eta^\prime}}$ can be generalized to the system with pairing terms
\begin{align}
\int d\mu(\boldsymbol{\xi})e^{\boldsymbol{\xi^*}\boldsymbol{\Theta}\boldsymbol{\xi}+\boldsymbol{\eta^*}\boldsymbol{\xi}+\boldsymbol{\xi^*}\boldsymbol{\eta^\prime}+\frac{1}{2}\boldsymbol{\xi^*}\boldsymbol{\mathcal{P}^\prime}\boldsymbol{\xi^\dagger}+\frac{1}{2}\boldsymbol{\xi^T}\boldsymbol{\mathcal{P}^*}\boldsymbol{\xi}}
=&\left\Vert\boldsymbol{1}-\boldsymbol{\Theta}\right\Vert^{-1/2}\left\Vert\boldsymbol{1}-\boldsymbol{\Phi}\right\Vert^{-1/2}\exp\left[\boldsymbol{\eta^*}(\boldsymbol{1}-\boldsymbol{\Phi})^{-1}\boldsymbol{\eta^\prime}\right]\notag\\
&\times \exp\left[\frac{1}{2}\boldsymbol{\eta^*}(\boldsymbol{1}-\boldsymbol{\Phi})^{-1}\boldsymbol{\mathcal{P}^\prime}(\boldsymbol{1}-\boldsymbol{\Theta^T})^{-1}\boldsymbol{\eta^\dagger}\right]\notag\\
&\times \exp\left[\frac{1}{2}\boldsymbol{\eta^{\prime T}}(\boldsymbol{1}-\boldsymbol{\Theta^T})^{-1}\boldsymbol{\mathcal{P}^*}(\boldsymbol{1}-\boldsymbol{\Phi})^{-1}\boldsymbol{\eta^\prime}\right]
\end{align}
where $\boldsymbol{\Phi}=\boldsymbol{\Theta}+\boldsymbol{\mathcal{P}^\prime}(\boldsymbol{1}-\boldsymbol{\Theta^T})^{-1}\boldsymbol{\mathcal{P}^*}$. By rewriting it in matrix form, the above equation can be reduced as 
\begin{align}
&\int d\mu(\boldsymbol{\xi})\exp\left[\frac{1}{2}\begin{pmatrix}
\boldsymbol{\xi^*} & \boldsymbol{\xi^T}
\end{pmatrix}
\begin{pmatrix}
\boldsymbol{\Theta}&\boldsymbol{\mathcal{P}^\prime}\\
\boldsymbol{\mathcal{P}^*}&\boldsymbol{\Theta^\dagger}
\end{pmatrix}
\begin{pmatrix}
\boldsymbol{\xi}\\
\boldsymbol{\xi^\dagger}
\end{pmatrix}
+\begin{pmatrix}
\boldsymbol{\xi^*}&\boldsymbol{\xi^T}
\end{pmatrix}
\begin{pmatrix}
\boldsymbol{\eta^\prime}\\
\boldsymbol{\eta^\dagger}
\end{pmatrix}
\right]\notag\\
&=\left\Vert\boldsymbol{1}-\boldsymbol{\Theta}\right\Vert^{-1/2}\left\Vert\boldsymbol{1}-\boldsymbol{\Phi}\right\Vert^{-1/2}\exp\left[\frac{1}{2}\begin{pmatrix}
\boldsymbol{\eta^*} & \boldsymbol{\eta^{\prime T}}\end{pmatrix}
\begin{pmatrix}
(\boldsymbol{1}-\boldsymbol{\Phi})^{-1} & (\boldsymbol{1}-\boldsymbol{\Phi})\boldsymbol{\mathcal{P}}(\boldsymbol{1}-\boldsymbol{\Theta^\dagger})^{-1}\\
(\boldsymbol{1}-\boldsymbol{\Theta^\dagger})^{-1}\boldsymbol{\mathcal{P}^*}(\boldsymbol{1}-\boldsymbol{\Phi})^{-1} & (\boldsymbol{1}-\boldsymbol{\Phi^T})^{-1}
\end{pmatrix}
\begin{pmatrix}
\boldsymbol{\eta^\prime}\\
\boldsymbol{\eta^\dagger}
\end{pmatrix}
\right]\notag\\
&=\left\Vert\boldsymbol{1}-\begin{pmatrix}
\boldsymbol{\Theta}&\boldsymbol{\mathcal{P}^\prime}\\
\boldsymbol{\mathcal{P}^*}&\boldsymbol{\Theta^\dagger}
\end{pmatrix}\right\Vert^{-1/2}\exp\left\{\frac{1}{2}\begin{pmatrix}
\boldsymbol{\eta^*} & \boldsymbol{\eta^{\prime T}}\end{pmatrix}
\left[\boldsymbol{1}-\begin{pmatrix}
\boldsymbol{\Theta}&\boldsymbol{\mathcal{P}^\prime}\\
\boldsymbol{\mathcal{P}^*}&\boldsymbol{\Theta^\dagger}
\end{pmatrix}
\right]^{-1}
\begin{pmatrix}
\boldsymbol{\eta^\prime}\\
\boldsymbol{\eta^\dagger}
\end{pmatrix}
\right\}
\end{align}
By applying this generalized Gaussian integral to the integral in Eq.~(\ref{rhos_coherent}), we obtain
\begin{align}
\label{A8}
\hspace*{-9mm}\langle z_S|\rho_S
|z^\prime_S\rangle=&\dfrac{\sqrt{\det \bm{\Omega}}}{Z_{\rm{tot}} }\left\Vert\boldsymbol{1}-\begin{pmatrix}
\boldsymbol{\Omega_{EE}}  & \boldsymbol{\Pi_{EE}} \\
\boldsymbol{\Pi_{EE}^*}  & \boldsymbol{\Omega^*_{EE}} 
\end{pmatrix}\right\Vert^{-1/2}\exp\left[\dfrac{1}{2}\begin{pmatrix}
z_S^* & z_S^{\prime}
\end{pmatrix}
\begin{pmatrix}
\Omega_{SS}  & \Pi_{SS} \\
\Pi_{SS}^*  & \Omega_{SS}
\end{pmatrix}
\begin{pmatrix}
z_{S}^\prime\\
z_S^*
\end{pmatrix}
\right]\notag\\
&\times\exp\left\{\dfrac{1}{2}\begin{pmatrix}
z_S^* & z_S^{\prime}
\end{pmatrix}
\begin{pmatrix}
\boldsymbol{\Omega}_{S\bm{E}}  & \boldsymbol{\Pi}_{S\bm{E}} \\
\boldsymbol{\Pi}_{S\bm{E}}^*  & \boldsymbol{\Omega}^*_{S\bm{E}} 
\end{pmatrix}
\left[\boldsymbol{1}-\begin{pmatrix}
\boldsymbol{\Omega_{EE}}  & \boldsymbol{\Pi_{EE}} \\
\boldsymbol{\Pi_{EE}^*}  & \boldsymbol{\Omega^*_{EE}} 
\end{pmatrix}\right]^{-1}
\begin{pmatrix}
\boldsymbol{\Omega}_{\bm{E}S}  & \boldsymbol{\Pi}_{\bm{E}S} \\
\boldsymbol{\Pi}_{\bm{E}S}^*  & \boldsymbol{\Omega}^*_{\bm{E}S} 
\end{pmatrix}
\begin{pmatrix}
z_{S}^\prime\\
z_S^*
\end{pmatrix}
\right\}
\notag\\
=&\dfrac{1}{Z^r_{S_0} }\exp\left[\dfrac{1}{2}\begin{pmatrix}
z_S^* & z_S^{\prime}
\end{pmatrix}
\begin{pmatrix}
\Omega_{S}  & \Pi_{S} \\
\Pi_{S}^*  & \Omega_{S}
\end{pmatrix}
\begin{pmatrix}
z_{S}^\prime\\
z_S^*
\end{pmatrix}
\right].
\end{align}
This is the result of Eq.~(\ref{coherent_S}). Furthermore, the correlation functions $n= \Tr[a^\dag a \rho_S]$ and $s= \Tr[aa \rho_S]$ can also easily be calculated in terms of the Gaussian kernel element by using the same generalized Gaussian integral in the coherent state presentation,
\begin{align}
\begin{pmatrix}
n &s \\
s^* &h
\end{pmatrix}&\equiv\begin{pmatrix}
\langle a^\dagger a \rangle & \langle a a \rangle\\
\langle a^\dagger a^\dagger \rangle & \langle a a^\dagger \rangle
\end{pmatrix}\notag\\
&=\dfrac{1}{(1-\Omega_S)^2-|\Pi_S|^2}\begin{pmatrix}
\Omega_S(1-\Omega_S)+|\Pi_S|^2&
\Pi_S\\
\Pi_S^*&
1-\Omega_S
\end{pmatrix},
\end{align}
which gives the results of Eq.~(\ref{NS}).

In order to find the reduced density matrix in an operator form, we use the same technique of the faithful matrix representation of group theory \cite{group}. The above reduced density matrix in the coherent state representation can be rewritten as
\begin{align}
\label{rdmo}
\langle z_S\vert\rho_S\vert z_S^\prime\rangle=&\dfrac{1}{Z_{S_0}^r} e^{\frac{1}{2} \Pi_2(z^*_S)^2 + z^*_S \Omega_S z'_S
+ \frac{1}{2} \Pi^*_S (z'_S)^2} \notag\\
=&\dfrac{e^{-\gamma}}{Z_{S_0}^r}\langle z_S|e^{\alpha a^{\dagger2}}e^{\gamma(a^\dagger a+aa^\dagger)}e^{\alpha^* a^2}|z_S^\prime\rangle
\end{align}
Thus, we obtain 
\begin{align}
\label{A10}
\rho_S=\dfrac{1}{Z_S^r}e^{\alpha a^{\dagger2}}e^{\gamma(a^\dagger a+aa^\dagger)}e^{\alpha^* a^2},
\end{align}
where

\begin{subequations}
\begin{align}
&\alpha=\dfrac{\Pi_S}{2}=\dfrac{s}{2[(1+n)^2-|s|^2]}\\
&\gamma=\ln\sqrt{\Omega_S}=\ln\sqrt{1-\dfrac{1+n}{(1+n)^2-|s|^2}}\\
&Z_S^r=e^\gamma Z^r_{S_0}=\sqrt{\dfrac{\Omega_S}{(1-\Omega_S)^2-|\Pi_S|^2}}=\sqrt{n^2+n-|s|^2}
\end{align}
\end{subequations}

Furthermore, using the Baker-Campbell-Hausdroff formula of the group $H_6$ \cite{group}
, Eq.~(\ref{rdmo}) can be reexpressed as 
\begin{align}
\rho_S=\dfrac{1}{Z_S^r}\exp\left[\eta(a^\dagger a+aa^\dagger)+\delta^* a^{\dagger2}+\delta a^2\right].
\end{align}
Using the faithful matrix representation again, we can determine the coefficients $\eta$ and $\delta$:
\begin{align}
\dfrac{1}{Z_S^r}e^{\eta(a^\dagger a+aa^\dagger)+\delta^* a^{\dagger2}+\delta a^2}=\begin{pmatrix}
1&0&0&0\\
0&\cosh\theta+\frac{2\eta}{\theta}\sinh\theta&\frac{2\delta^*}{\theta}\sinh\theta&0\\
0&-\frac{2\delta}{\theta}\sinh\theta&\cosh\theta-\frac{2\eta}{\theta}\sinh\theta&0\\
2\ln Z_S^r&0&0&1
\end{pmatrix},
\end{align}
where $\theta=2\sqrt{\left\vert\eta^2-|\delta|^2\right\vert}$. Comparing it to the faithful matrix representation of the reduced density matrix
\begin{align}
\dfrac{1}{Z_S^r}e^{\alpha a^{\dagger2}}e^{\gamma(a^\dagger a+aa^\dagger)}e^{\alpha^* a^2}&=\begin{pmatrix}
1&0&0&0\\
0&\Omega_S-|\Pi_S|^2\Omega_S^{-1}&\Pi_S\Omega_S^{-1}&0\\
0&-\Pi^*_S\Omega_S^{-1}&\Omega_S^{-1}&0\\
2\ln Z_S^r&0&0&1
\end{pmatrix}\notag\\
& =\begin{pmatrix}
1&0&0&0\\
0&\dfrac{n^2-|s|^2}{n^2+n-|s|^2}&\dfrac{s}{n^2+n-|s|^2}&0\\
0&-\dfrac{s^*}{n^2+n-|s|^2}&\dfrac{(1+n)^2-|s|^2}{n^2+n-|s|^2}&0\\
2\ln Z_S^r&0&0&1
\end{pmatrix},
\end{align}
we obtain the relations presented in Eq.~(\ref{12}).

\section{Derivation of the reduced density matrix using coherent state path integral}
\label{appendixB}
In this Appendix, we compute the reduced density matrix with Euclidean-space path integral in the coherent state representation. Begin with the total-equilibrium density matrix $\rho_{\rm{tot}}=\frac{1}{Z_{\rm{tot}}}e^{-\beta H_{\rm{tot}}}$ in Eq.~(\ref{tGS}) and rewrite $\beta$ as $\beta=\frac{1}{\hbar}(\tau_f-\tau_i)$, we have the reduced density matrix in the coherent state representation,
\begin{align}
\langle z_{S,f}|\rho_{\rm{S}}|z_{S,i}\rangle=\dfrac{1}{Z_{\rm{tot}}}\int d\mu(\bm{z}_{\bm{E},f}) \langle z_{S,f},\bm{z}_{\bm{E},f}|U(\tau_f,\tau_i)|z_{S,i},\bm{z}_{\bm{E},f}\rangle
\end{align}
where $U(\tau_f,\tau_i)=e^{-\frac{1}{\hbar}H_{\rm{tot}}(\tau_f-\tau_i)}$ is the imaginary-time evolution operator, and the evolution along the imaginary time $\tau$ is periodic with the period $\hbar\beta$, that is, $\bm{z}(\tau)=\bm{z}(\tau+\hbar\beta)$. Then we divide the imaginary-time interval into small pieces $\delta\beta=\beta/N\rightarrow0$
\begin{align}
&\int d\mu(\bm{z}_{\bm{E},f}) \langle z_{S,f},\bm{z}_{\bm{E},f}|U(\tau_f,\tau_i)|z_{S,i},\bm{z}_{\bm{E},f}\rangle= \int d\mu(\bm{z}_{\bm{E},f}) \langle z_{S,f},\bm{z}_{\bm{E},f}|(e^{-\delta \beta H_{\rm{tot}}})^N|z_{S,i},\bm{z}_{\bm{E},f}\rangle\notag\\
&=\prod\limits_{n=1}^{N-1}\int d\mu(z_{S,n}) \prod\limits_{n=1}^N d\mu(\bm{z}_{\bm{E},n}) \langle z_{S,n},\bm{z}_{\bm{E},n}|e^{-\delta \beta H_{\rm{tot}}}|z_{S,n-1},\bm{z}_{\bm{E},n-1}\rangle\notag\\
&=
\prod\limits_{n=1}^{N-1}\int d\mu(z_{S,n}) \prod\limits_{n=1}^N d\mu(\bm{z}_{\bm{E},n}) \langle z_{S,n},\bm{z}_{\bm{E},n}|z_{S,n-1},\bm{z}_{\bm{E},n-1}\rangle\left(1-\delta\beta\dfrac{\langle z_{S,n},\bm{z}_{\bm{E},n}|H_{\rm{tot}}|z_{S,n-1},\bm{z}_{\bm{E},n-1}\rangle}{\langle z_{S,n},\bm{z}_{\bm{E},n}|z_{S,n-1},\bm{z}_{\bm{E},n-1}\rangle}\right)\notag\\
&=\prod\limits_{n=1}^{N-1}\int d\mu(z_{S,n})\prod\limits_{n=1}^N\!\exp\!\left[z^*_{S,n}z_{S,n-1}+\bm{z}^\dag_{\bm{E},n-1}\bm{z}_{\bm{E},n-1}\right]\notag\\
&\hspace{10mm}\times\prod\limits_{n=1}^N d\mu(\bm{z}_{\bm{E},n})\!\exp\!\left\{-\delta\beta[H_S(z^*_{S,n},z_{S,n-1})+H_E(\bm{z}^*_{\bm{E},n-1},\bm{z}_{\bm{E},n-1})+H_I(z^*_{S,n},z_{S,n-1},\bm{z}^*_{\bm{E},n-1},\bm{z}_{\bm{E},n-1})]\right\}\notag\\
&=\int {\cal D}\mu[z_S(\tau)]e^{-\frac{1}{\hbar}S_s[z_S^*,z_S]}\mathcal{F}(z_{S,f}^*,\tau_f;z_{S,i},\tau_i),
\end{align}
where $z^*_{S,N}=z^*_{S,f}, \ z_{S,0}=z_{S,i}, \ \bm{z}^*_{\bm{E},N}=\bm{z}^*_{\bm{E},f}, \ \bm{z}_{\bm{E},0}=\bm{z}_{\bm{E},f}$, and $S_s[z_S^*,z_S]$ is the Euclidean action of system, the action in the Euclidean spacetime obtained by performing Wick Rotation on the Minkowski action, which is given by
\begin{align}
S_s[z_S^*,z_S]=\dfrac{\hbar}{2}[z_{S,f}^*z(\tau_f)+z^*(\tau_i)z_{S,i}]+\int_{\tau_i}^{\tau_f}d\tau\left\{\dfrac{\hbar}{2}[\dot{z}^*(\tau)z_S(\tau)-z_S^*(\tau)\dot{z}(\tau)]-H_S(z_S^*(\tau),z_S(\tau))\right\}
\end{align}
and the remaining part of propagator is the influence functional arisen from the partial trace over all the states of the reservoir
\begin{align}
&\mathcal{F}(z_{S,f}^*,\tau_f;z_{S,i},\tau_i)=\int D\mu[{\bm{z}_{\bm{E}}(\tau)}]\exp\frac{1}{\hbar}\left\{-S_E[\bm{z_E}^*,\bm{z_E}]-S_I[z_S^*,z_S,\bm{z_E}^*,\bm{z_E}]\right\} 
\label{influence functional}
\end{align}
including the end-point integral $\int d\mu(\bm{z}_{\bm{E},f})$, where the Euclidean-space actions of the reservoir and the system-reservoir interaction are
\begin{align}
&S_E[\bm{z_E}^*,\bm{z_E}]=\dfrac{\hbar}{2}[\bm{z}_{\bm{E},f}^\dag\bm{z_E}(\tau_f)+\bm{z_E}^\dag(\tau_i)\bm{z}_{\bm{E},i}]+\int_{\tau_i}^{\tau_f}d\tau\left\{\dfrac{\hbar}{2}[\dot{\bm{z}}_{\bm{E}}^\dag(\tau)\bm{z}_{\bm{E}}(\tau)-\bm{z_E}^\dag(\tau)\dot{\bm{z}}_{\bm{E}}(\tau)]- H_E(\bm{z_E}^*(\tau),\bm{z}_{\bm{E}}(\tau))\right\}\notag\\
&S_I[z_S^*,z_S,\bm{z_E}^*,\bm{z_E}]=-\int_{\tau_i}^{\tau_f}d\tau H_I(z^*,z,\bm{z_E}^*(\tau),\bm{z}_{\bm{E}}(\tau)).
\end{align}
The quadratic form of these actions allows the path integral of influence functional of Eq.~(\ref{influence functional}) to be solved exactly with the stationary path. The stationary path of the Euclidean action obeys the Wick-rotated version of the equations of motion 
\begin{subequations}
\begin{align}
&\left\{\begin{array}{l}
\dot{z}_{E_k}(\tau)\!=\!-\omega_kz_{E_k}(\tau)- V_k[z_S(\tau)\!+\!z_S^*(\tau)]\\[+2mm]
\dot{z}_{E_k}^*(\tau)\!=\!\omega_kz_{E_k}^*(\tau)+ V^*_k[z_S(\tau)\!+\!z_S^*(\tau)]
\end{array}\right.\!
\Rightarrow
\left\{\begin{array}{l}
z_{E_k}(\tau)\!=\!z_{E_k,i}e^{-\omega_k(\tau-\tau_i)}\!-\! V_k\int_{\tau_i}^{\tau}d\tau'e^{-\omega_k(\tau-\tau')}[z_S(\tau')\!+\!z_S^*(\tau')]\\[+2mm]
z_{E_k}^*(\tau)\!=\!z^*_{E_k,f}e^{-\omega_k(\tau_f-\tau)}\!-\! V^*_k\int_{\tau}^{\tau_f}d\tau'e^{\omega_k(\tau-\tau')}[z_S^*(\tau')\!+\!z_S(\tau')]\end{array}\right.
\end{align}
\end{subequations}
Substitute these solutions into $S_E[\bm{z_E}^*,\bm{z_E}]$ and $S_I[z_S^*,z_S,\bm{z_E}^*,\bm{z_E}]$, we have
\begin{align}
S_E[\bm{z_E}^*,\bm{z_E}]+S_I[z_S^*,z_S,\bm{z_E}^*,\bm{z_E}]=&\dfrac{\hbar}{2}[\bm{z}_{\bm{E},f}^\dag\bm{z_E}(\tau_f)+\bm{z_E}^\dag(\tau_i)\bm{z}_{\bm{E},i}]+\hbar\int_{\tau_i}^{\tau_f}d\tau\bigg\{\frac{1}{2}[\dot{\bm{z_E}}^\dag(\tau)\bm{z}_{\bm{E}}(\tau)-\bm{z_E}^\dag(\tau)\dot{\bm{z_E}}(\tau)]\notag\\
&-\!\sum\limits_k\!\omega_kz_{E_k}^*(\tau)z_{E_k}(\tau)\!+\!V_kz_{E_k}^*(\tau)[z_S(\tau)\!+\!z_S^*(\tau)]\!+\!V^*_k[z_S(\tau)\!+\!z_S^*(\tau)]z_{E_k}(\tau)\bigg\}\notag\\
=&\hbar\sum\limits_kz_{E_k,f}^*z_{E_k,i}e^{-\omega_k(\tau_f-\tau_i)}\notag\\
&-\hbar\sum\limits_kV_k\int_{\tau_i}^{\tau_f}d\tau e^{-\omega_k(\tau_f-\tau)}z^*_{E_k,f}[z_S(\tau)+z_S^*(\tau)]\notag\\
&-\hbar\sum\limits_kV^*_k\int_{\tau_i}^{\tau_f}d\tau e^{-\omega_k(\tau-\tau_i)}[z_S(\tau)+z_S^*(\tau)]z_{E_k,i}\notag\\
&+\hbar\sum\limits_k|V_k|^2\int_{\tau_i}^{\tau_f}\!d\tau\int_{\tau_i}^{\tau}d\tau'[z_S(\tau)+z_S^*(\tau)]e^{-\omega_k(\tau-\tau')}\theta(\tau-\tau')[z_S(\tau)+z_S^*(\tau)].
\end{align}
Then the influence functional of Eq.~(\ref{influence functional}) can be expressed as
\begin{align}
&\mathcal{F}(z_{S,f}^*,\tau_f;z_{S,i},\tau_i)\notag\\
&=\int d\mu(\bm{z}_{\bm{E},f}) \prod\limits_k\exp\left[\begin{array}{l}
z_{E_k,f}^* U_{E_k}(\tau_f,\tau_i)z_{E_k,f}\\[+2mm]
-\int_{\tau_i}^{\tau_f}d\tau
[z_S^*(\tau)+z_S(\tau)]V^*_k
U_{E_k}(\tau,\tau_i)
z_{E_k,f}\\[+2mm]
-\int_{\tau_i}^{\tau_f}d\tau
z^*_{E_k,f}
U_{E_k}(\tau_f,\tau')
V_k[z_S^*(\tau')+z_S(\tau')]\\[+2mm]
+\int_{\tau_i}^{\tau_f}d\tau\int_{\tau_i}^{\tau}d\tau'
[z_S^*(\tau)+z_S(\tau)]V^*_k
U_{E_k}(\tau-\tau')\theta(\tau-\tau')
V_k[z_S^*(\tau')+z_S(\tau')]
\end{array}\right],
\end{align}
where $U_{E_k}(\tau-\tau')=e^{-\omega_k(\tau-\tau')}$. The environment variable can be trace out through Gaussian integral
\begin{align}
&\mathcal{F}(z_{S,f}^*,\tau_f;z_{S,i},\tau_i)\notag\\
&=\prod\limits_{k,k'}(1-e^{-\beta\hbar\omega_{k'}})\exp\left[
\begin{array}{l}
\int_{\tau_i}^{\tau_f}d\tau\int_{\tau_i}^{\tau}d\tau'[z_S^*(\tau)+z_S(\tau)]V_{k}^*
U_{E_k}(\tau,\tau_i)(
1-e^{-\beta\hbar\omega_k})^{-1}
U_{E_k}(\tau_f,\tau')
V_k[z_S^*(\tau')+z_S(\tau')]\\[+2mm]
+\int_{\tau_i}^{\tau_f}d\tau\int_{\tau_i}^{\tau}d\tau'
[z_S^*(\tau)+z_S(\tau)]V_k^*
U_{E_k}(\tau-\tau')\theta(\tau-\tau')
V_k[z_S^*(\tau')+z_S(\tau')]
\end{array}\right]\notag\\
&=\prod\limits_k(1-e^{-\beta\hbar\omega_k})\exp\left[\int_{\tau_i}^{\tau_f}d\tau\int_{\tau_i}^{\tau}d\tau'[z_S^*(\tau)+z_S(\tau)]
[g(\tau-\tau')\theta(\tau-\tau')+g'(\tau-\tau')]
[z_S(\tau')+z_S^*(\tau')]\right],
\end{align}
where the integral kernels are given by
\begin{subequations}
\begin{align}
g(\tau-\tau')&=\sum\limits_k|V_k|^2e^{-\omega_k(\tau-\tau')}=\int\dfrac{d\omega}{2\pi}J(\omega)e^{-\omega(\tau-\tau')}\\
g'(\tau-\tau')&=\sum\limits_k|V_k|^2e^{-\omega_k(\tau-\tau')}e^{-\beta\hbar\omega_k}/(1-e^{-\beta\hbar\omega_k})=\int\dfrac{d\omega}{2\pi}J(\omega)e^{-\omega(\tau-\tau')}/(e^{\beta\hbar\omega}-1),
\end{align}
\end{subequations}
in which the spectral density is defined as
\begin{align}
J(\epsilon)=2\pi\sum\limits_k|V_k|^2\delta(\omega-\omega_k).
\end{align}
Then the reduced density matrix in the coherent state representation is
\begin{align}
\langle z_{S,f}|\rho_S|z_{S,i}\rangle=\dfrac{1}{Z_{\rm{tot}}}\prod\limits_k(1-e^{-\beta\hbar\omega_k})\int_{z_{S,i}}^{z_{S,f}}D\mu[z_S(\tau)]e^{-\frac{1}{\hbar}S_{\rm{eff}}[z_S^*,z_S]}
\end{align}
where the effective action is
\begin{align}
\frac{1}{\hbar}S_{\rm{eff}}=&\dfrac{1}{2}[z_{S,f}^*z_S(\tau_f)+z_S^*(\tau_i)z_{S,i}]+\int_{\tau_i}^{\tau_f}d\tau\bigg\{\dfrac{1}{2}[\dot{z}_S^*(\tau)z_S(\tau)-z_S^*(\tau)\dot{z}_S(\tau)]-z_S^*(\tau)\omega_Sz_S(\tau)\notag\\
&+[z_S^*(\tau)+z_S(\tau)][g(\tau-\tau')\theta(\tau-\tau')+g'(\tau-\tau')][z_S^*(\tau')+z_S(\tau')]\bigg\}.
\end{align}
Expanding the action by variation with respect to the stationary path, we have
\begin{align}
\frac{1}{\hbar}\delta S_{\rm{eff}}=&\dfrac{1}{2}\int_{\tau_i}^{\tau_f}d\tau[\dot{z}_S^*(\tau)\delta z_S(\tau)-z_S^*(\tau)\omega_S\delta z_S(\tau)]\notag\\
&+\dfrac{1}{2}\int_{\tau_i}^{\tau_f}d\tau\delta z_S(\tau)[g(\tau-\tau')\theta(\tau-\tau')+g'(\tau-\tau')][z_S^*(\tau')+z_S(\tau')]\notag\\
&+\dfrac{1}{2}\int_{\tau_i}^{\tau_f}d\tau[z_S^*(\tau')+z_S(\tau')][g(\tau'-\tau)\theta(\tau-\tau')+g'(\tau'-\tau)]\delta z_S(\tau)\notag\\
&+\dfrac{1}{2}\int_{\tau_i}^{\tau_f}d\tau[-\delta z_S^*(\tau)][\dot{z}_S(\tau)-\delta z_S^*(\tau)\omega_Sz_S(\tau)]\notag\\
&+\dfrac{1}{2}\int_{\tau_i}^{\tau_f}d\tau\delta z_S^*(\tau)[g(\tau-\tau')\theta(\tau-\tau')+g'(\tau-\tau')][z_S^*(\tau')+z_S(\tau')]\notag\\
&+\dfrac{1}{2}\int_{\tau_i}^{\tau_f}d\tau[z_S^*(\tau')+z_S(\tau')][g(\tau'-\tau)\theta(\tau-\tau')+g'(\tau'-\tau)]\delta z_S^*(\tau).
\label{A15}
\end{align}
Taking stationary path such that $\delta S_{\rm{eff}}|_{st} = 0$, we can obtain the equations of motion
\begin{align}
\dot{z}_S(\tau)=&-\omega_Sz_S(\tau)+\int_{\tau_i}^{\tau}d\tau'g(\tau-\tau')[z_S^*(\tau')+z_S(\tau')]-\int_{\tau}^{\tau_f}d\tau'g(\tau'-\tau)[z_S^*(\tau')+z_S(\tau')]\notag\\
&+\int_{\tau_i}^{\tau_f}d\tau'[g'(\tau'-\tau)+g'(\tau-\tau')][z_S^*(\tau')+z_S(\tau')]\notag\\
\dot{z}^*_S(\tau)=&\omega_Sz_S^*(\tau)-\int_{\tau_i}^{\tau}d\tau'g(\tau-\tau')[z_S^*(\tau')+z_S(\tau')]+\int_{\tau}^{\tau_f}d\tau'g(\tau'-\tau)[z_S^*(\tau')+z_S(\tau')]\notag\\
&-\int_{\tau_i}^{\tau_f}d\tau'[g'(\tau'-\tau)+g'(\tau-\tau')][z_S^*(\tau')+z_S(\tau')].
\end{align}
Then
\begin{align}
S_{\rm{eff}}[z_S^*,z_S]=&\dfrac{1}{2}[z_{S,f}^*z_S(\tau_f)+z_S^*(\tau_i)z_{S,i}]+\int_{\tau_i}^{\tau_f}d\tau z_S^*(\tau)\omega_Sz_S(\tau)\notag\\
&+\int_{\tau_i}^{\tau_f}d\tau [z_S^*(\tau)+z_S(\tau)][g(\tau-\tau')+g'(\tau-\tau')][z_S^*(\tau')+z_S(\tau')]\notag\\
&-\dfrac{1}{2}\int_{\tau_i}^{\tau_f}d\tau z_S^*(\tau)\left\{-\omega_Sz_S(\tau)+[g(\tau-\tau')+g'(\tau-\tau')][z_S^*(\tau')+z_S(\tau')]\right\}\notag\\
&-\dfrac{1}{2}\int_{\tau_i}^{\tau_f}d\tau\left\{-z_S^*(\tau)\omega_S+[z_S^*(\tau')+z_S(\tau')][g(\tau-\tau')+g'(\tau-\tau')]\right\}z_S(\tau)\notag\\
=&\dfrac{1}{2}[z_{S,f}^*z_S(\tau_f)+z_S^*(\tau_i)z_{S,i}]
\end{align}
To solve the integro-differential equations Eq.~(\ref{A15}), we make a transformation
\begin{align}
\begin{pmatrix}
z_S(\tau)\\z_S^*(\tau)
\end{pmatrix}=\begin{pmatrix}
\Omega_S(\tau-\tau_i)&\Pi_S(\tau-\tau_i)\\
\Pi_S^*(\tau_f-\tau)&\Omega_S^*(\tau_f-\tau)
\end{pmatrix}\begin{pmatrix}
z_{S,i}\\z_{S,f}^*
\end{pmatrix}.
\end{align}
Then, we have
\begin{align}
\langle z_{S,f}|\rho_S|z_{S,i}\rangle=&\dfrac{1}{Z_{\rm{tot}}}\prod\limits_k(1-e^{-\beta\hbar\omega_k})\int_{z_{S,i}}^{z^*_{S,f}}D\mu[z_S(\tau)]e^{-\frac{1}{\hbar}S_{\rm{eff}}[z_S^*,z_S]}\notag\\
=&\dfrac{1}{Z_{\rm{tot}}}\prod\limits_k(1-e^{-\beta\hbar\omega_k})K(\tau_f,\tau_i)
\exp\left[\dfrac{1}{2}\begin{pmatrix}
z_{S,f}^*&z_{S,i}
\end{pmatrix}\begin{pmatrix}
\Omega_S(\hbar\beta)&\Pi_S(\hbar\beta)\\
\Pi_S^*(\hbar\beta)&\Omega_S^*(\hbar\beta)
\end{pmatrix}
\begin{pmatrix}
z_{S,i}\\
z^*_{S,f}
\end{pmatrix}\right]
\end{align}
where $K(\tau_f,\tau_i)$ is a endpoint-independent constant which is given by
\begin{align}
K(\tau_f,\tau_i)=\int_{0}^{0^*}&D\mu[\delta z_S(\tau)]\exp\left\{\dfrac{1}{2}\int_{\tau_i}^{\tau_f}d\tau[\delta \dot{z}_S^*(\tau)\delta z_S(\tau)-\delta z_S^*(\tau)\delta \dot{z}_S(\tau)]-\int_{\tau_i}^{\tau_f}d\tau \delta z_S^*(\tau)\omega_S\delta z_S(\tau)
\right\}\notag\\
&\times\exp\left\{\int_{\tau_i}^{\tau_f}d\tau [\delta z_S^*(\tau)+\delta z_S(\tau)][g(\tau-\tau')\theta(\tau-\tau')+g'(\tau-\tau')][\delta z_S^*(\tau')+\delta z_S(\tau')]\right\},
\end{align}
and can be further determined by the condition $\int D\mu(z_S)\langle z_{S}|\rho_S|z_{S}\rangle=1$.

Now, the matrix elements are solvable at the point $\tau_f-\tau_i=\hbar\beta$ and are given by
\begin{align}
\dot{\Omega}_S(\tau)|_{\tau\rightarrow\hbar\beta}=&-\omega_S \Omega_S(\hbar\beta)+\int_{0}^{\hbar\beta}d\tau'\left\{ g(\hbar\beta-\tau')+[g'(\tau'-\hbar\beta)+g'(\hbar\beta-\tau')]\right\}[\Omega_S(\tau')+\Pi_S^*(\tau')]\notag\\
\dot{\Pi}^*_S(\tau)|_{\tau\rightarrow\hbar\beta}=&-\omega_S \Pi^*_S(\hbar\beta)+\int_{0}^{\hbar\beta}d\tau'\left\{ g(\hbar\beta-\tau')+[g'(\tau'-\hbar\beta)+g'(\hbar\beta-\tau')]\right\}[\Omega^*_S(\tau')+\Pi_S(\tau')]\notag\\
\dot{\Omega}^*_S(\tau)|_{\tau\rightarrow\hbar\beta}=&\omega_S \Omega^*_S(\hbar\beta)+\int_{0}^{\hbar\beta}d\tau'\left\{ g(\hbar\beta-\tau')-[g'(\tau'-\hbar\beta)+g'(\hbar\beta-\tau')]\right\}[\Omega^*_S(\tau')+\Pi_S(\tau')]\notag\\
\dot{\Pi}^*_S(\tau)|_{\tau\rightarrow\hbar\beta}=&\omega_S \Pi_S(\hbar\beta)+ \int_{0}^{\hbar\beta}d\tau'\left\{g(\tau'-\hbar\beta)-[g'(\tau'-\hbar\beta)+g'(\hbar\beta-\tau')]\right\}[\Omega_S(\tau')+\Pi_S^*(\tau')]
\end{align}
$\Rightarrow$
\begin{align}
\dfrac{d}{d\tau}\left.\begin{pmatrix}
\Omega_S(\tau)&\Pi_S(\tau)\\
\Pi_S^*(\tau)&\Omega_S^*(\tau)
\end{pmatrix}\right\vert_{\tau\rightarrow\hbar\beta}=&-\omega_S\begin{pmatrix}
1&0\\0&-1
\end{pmatrix}\begin{pmatrix}
\Omega_S(\hbar\beta)&\Pi_S(\hbar\beta)\\
\Pi_S^*(\hbar\beta)&\Omega_S^*(\hbar\beta)
\end{pmatrix}+\int_0^{\hbar\beta} d\tau' \begin{pmatrix}
g(\hbar\beta-\tau')&g(\hbar\beta-\tau')\\
g(\tau'-\hbar\beta)&g(\tau'-\hbar\beta)
\end{pmatrix}\begin{pmatrix}
\Omega_S(\tau')&\Pi_S(\tau')\\
\Pi_S^*(\tau')&\Omega_S^*(\tau')
\end{pmatrix}\notag\\
&+\int_0^{\hbar\beta} d\tau[g'(\tau-\hbar\beta)+g'(\hbar\beta-\tau)]\begin{pmatrix}
1&1\\-1&-1
\end{pmatrix}
\begin{pmatrix}
\Omega_S(\tau)&\Pi_S(\tau)\\
\Pi_S^*(\tau)&\Omega_S^*(\tau)
\end{pmatrix},
\end{align}
where the matrix elements are subjected to the initial conditions $\Omega_S(\tau_i)=1$ and $\Pi_S(\tau_i)=0$. Take the Laplace transform from $\tau$-domain to $s$-domain
\begin{align}
s\begin{pmatrix}
\tilde{\Omega}_S(s)&\tilde{\Pi}_S(s)\\
\tilde{\Pi}_S^*(s)&\tilde{\Omega}_S^*(s)
\end{pmatrix}-\begin{pmatrix}
1&0\\0&1
\end{pmatrix}=&-\omega_S\begin{pmatrix}
1&0\\0&-1
\end{pmatrix}\begin{pmatrix}
\tilde{\Omega}_S(s)&\tilde{\Pi}_S(s)\\
\tilde{\Pi}_S^*(s)&\tilde{\Omega}_S^*(s)
\end{pmatrix}\notag\\
&+\begin{pmatrix}
\Sigma(s)+\Sigma'(s)+\Sigma'(-s)&\Sigma(s)+\Sigma'(s)+\Sigma'(-s)\\
\Sigma(-s)-\Sigma'(s)-\Sigma'(-s)&\Sigma(-s)-\Sigma'(s)-\Sigma'(-s)
\end{pmatrix}\begin{pmatrix}
\tilde{\Omega}_S(s)&\tilde{\Pi}_S(s)\\
\tilde{\Pi}_S^*(s)&\tilde{\Omega}_S^*(s)
\end{pmatrix}\notag\\
&\hspace{-50mm}\Rightarrow\begin{pmatrix}
\tilde{\Omega}_S(s)&\tilde{\Pi}_S(s)\\
\tilde{\Pi}_S^*(s)&\tilde{\Omega}_S^*(s)
\end{pmatrix}=\begin{pmatrix}
s+\omega_S+\Sigma(s)+\Sigma'(s)+\Sigma'(-s)&\Sigma(s)+\Sigma'(s)+\Sigma'(-s)\\
\Sigma(-s)-\Sigma'(s)-\Sigma'(-s)&s-\omega_S+\Sigma(-s)-\Sigma'(s)-\Sigma'(-s)
\end{pmatrix}^{-1}
\end{align}
where $\Sigma(s)=\mathcal{L}[g(\tau)]$, $\Sigma'(s)=\mathcal{L}[g'(\tau)]$. Again, take the inverse Laplace transform reverts to the original domain and substitute $\tau=\hbar\beta$, then we have the solutions of $\Omega_S(\hbar\beta)$ and $\Pi_S(\hbar\beta)$ as given by Eq.~(\ref{OPS_Laplace}).
\end{widetext}

\end{document}